\documentclass[lettersize,journal]{IEEEtran}
\usepackage{amsmath,amsfonts, amssymb}
\usepackage{algorithmic}
\usepackage{algorithm}
\usepackage{array}
\usepackage[caption=false,font=footnotesize,labelfont=footnotesize,textfont=footnotesize]{subfig}
\usepackage{textcomp}
\usepackage{multirow}
\usepackage{threeparttable}
\usepackage{bbding}
\usepackage{makecell}
\usepackage{stfloats}
\usepackage{url}
\usepackage{verbatim}
\usepackage{graphicx}
\usepackage{cite}
\begin{document}

\title{Fundamental Detection Probability vs. Achievable Rate Tradeoff in Integrated Sensing and Communication Systems}
\author{Jiancheng An,~\IEEEmembership{Member,~IEEE},
Hongbin Li, \IEEEmembership{Fellow,~IEEE},
Derrick Wing Kwan Ng, \IEEEmembership{Fellow,~IEEE},\\
and Chau Yuen, \IEEEmembership{Fellow,~IEEE}
\thanks{This research is supported by the Ministry of Education, Singapore, under its MOE Tier 2 (Award number MOE-T2EP50220-0019). Any opinions, findings and conclusions or recommendations expressed in this material are those of the author(s) and do not reflect the views of the Ministry of Education, Singapore. This research is supported by A*STAR under its RIE2020 Advanced Manufacturing and Engineering (AME) Industry Alignment Fund – Pre Positioning (IAF-PP) (Grant No. A19D6a0053). Any opinions, findings and conclusions or recommendations expressed in this material are those of the author(s) and do not reflect the views of A*STAR. The work of H. Li was supported in part by the National Science Foundation under Grants ECCS-1923739 and ECCS-2212940. D. W. K. Ng is supported by the Australian Research Council's Discovery Project (DP210102169, DP230100603).}
\thanks{J. An is with the Engineering Product Development (EPD) Pillar, Singapore University of Technology and Design, Singapore 487372 (e-mail: jiancheng\_an@sutd.edu.sg). H. Li is with the Department of Electrical and Computer Engineering, Stevens Institute of Technology, Hoboken, NJ 07030, USA (e-mail: hli@stevens.edu). D. W. K. Ng is with the School of Electrical Engineering and Telecommunications, University of New South Wales (UNSW), Sydney, NSW 2052, Australia (e-mail: w.k.ng@unsw.edu.au). C. Yuen is with the School of Electrical and Electronics Engineering, Nanyang Technological University, Singapore 639798 (e-mail: chau.yuen@ntu.edu.sg).}}

\maketitle
\begin{abstract}
Integrating sensing functionalities is envisioned as a distinguishing feature of next-generation mobile networks, which has given rise to the development of a novel enabling technology -- \emph{Integrated Sensing and Communication (ISAC)}. Portraying the theoretical performance bounds of ISAC systems is fundamentally important to understand how sensing and communication functionalities interact (e.g., competitively or cooperatively) in terms of resource utilization, while revealing insights and guidelines for the development of effective physical-layer techniques. In this paper, we characterize the fundamental performance tradeoff between the detection probability for target monitoring and the user's achievable rate in ISAC systems. To this end, we first discuss the achievable rate of the user under sensing-free and sensing-interfered communication scenarios. Furthermore, we derive closed-form expressions for the probability of false alarm (PFA) and the successful probability of detection (PD) for monitoring the target of interest, where we consider both communication-assisted and communication-interfered sensing scenarios. In addition, the effects of the unknown channel coefficient are also taken into account in our theoretical analysis. Based on our analytical results, we then carry out a comprehensive assessment of the performance tradeoff between sensing and communication functionalities. Specifically, we formulate a power allocation problem to minimize the transmit power at the base station (BS) under the constraints of ensuring a required PD for perception as well as the communication user's quality of service requirement in terms of achievable rate. It indicates that, on the one hand, there exists an intrinsic tradeoff between sensing and communication performance under the mutual-interfered scenarios; On the other hand, with prior knowledge of the baseband waveform, these two functionalities might mutually assist each other to enhance the performance. Finally, simulation results corroborate the accuracy of our theoretical analysis and the effectiveness of the proposed power allocation solutions showing the advantages of the ISAC system over the conventional radar and communication coexistence counterpart.
\end{abstract}


\section{Introduction}
\IEEEPARstart{N}{ext-generation} radio access networks (RANs) are envisioned as a pivotal enabler to support various emerging environment-aware applications, which generally demand massive wireless connectivity as well as high-accuracy and robust sensing capabilities \cite{VTM_2021_Leyva_Cooperative}, such as simultaneous localization and mapping for autonomous driving \cite{WC_2022_An_Codebook}, Wi-Fi sensing for home health monitoring \cite{IOTJ_2020_Liu_Collaborative}, and super-definition imaging for extended reality \cite{Network_2020_Saad_A}. Fortunately, with the widespread deployment of the fifth-generation (5G) nodes enabled by millimeter-wave and massive multiple-input multiple-output (MIMO) technologies \cite{TWC_2023_Xu_Antenna, CST_2017_Busari_Millimeter}, future mobile networks are capable of providing high resolution in both the time and spatial domain, rendering its possibility in achieving high-accuracy perception by utilizing ubiquitous communication signals radiated from existing ultra-dense cellular infrastructure \cite{TVT_2022_Xu_Reconfigurable}. As a result, it is natural to amalgamate sensing and communication functionalities in beyond-5G/6G networks, which has motivated the recent research upsurge of \emph{Integrated Sensing and Communications (ISAC)} \cite{JCS_2021_Tan_Integrated}.

Historically, radar sensing and wireless communication technologies have been developed separately and have often competed with each other over the limited radio resources \cite{TCOM_2020_Liu_Joint, SPM_2019_Zheng_Radar}. Nevertheless, the essential pursuit for augmented performance is driving these two independent developments to interact with each other and ultimately evolve towards the regime of high-frequency band and large-scale antenna array regime \cite{SPM_2007_Li_MIMO, SPM_2019_Mishra_Toward, SPM_2012_Rusek_Scaling}. Motivated by this observation, ISAC, where sensing and communication functionalities share the same frequency band and hardware, is expected to substantially improve the spectral and energy efficiency, while reducing both hardware cost and signaling overheads \cite{CST_2022_Liu_A, Network_2021_Cui_Integrating}. In particular, ISAC pursues mutualistic integration of these two embedded functionalities to strike favorable tradeoffs between them and attain desired performance gains \cite{JSAC_2022_Liu_Integrated}. Specifically, on the one hand, the communication waveform in an ISAC system can be exploited for sensing the surrounding environment, e.g., buildings, pedestrians, vehicular traffic, etc. On the other hand, sensory data, such as position, angle-of-arrival, user's speed, etc., can be utilized to enhance the communication quality-of-service (QoS) in return.

Motivated by the aforementioned benefits, ISAC has recently attracted tremendous research interests from both academia and industry \cite{TWC_2021_Qi_Integrated, CL_2022_Zhao_Joint, JSAC_2022_Xiao_Waveform, TVT_2022_Zhang_Integrated, JSTSP_2021_Zhang_Design, TGCN_2022_An_Joint}. For example, the authors of \cite{TWC_2021_Qi_Integrated} presented a comprehensive ISAC framework for massive Internet-of-Things (IoT) systems, in which a pair of joint beamformers were proposed to manage the co-channel interference, thereby maximizing the weighted sum rate. Furthermore, a joint design of transmit and receive beamforming vectors for general ISAC systems was considered in \cite{CL_2022_Zhao_Joint} for maximizing the signal-to-interference-plus-noise-ratio (SINR) at the receiver, where the channel uncertainty was taken into account. Also, in \cite{JSAC_2022_Xiao_Waveform}, the authors proposed a novel full-duplex (FD) ISAC scheme by transmitting information-bearing signals during the waiting interval of conventional pulsed radars, which significantly increases the achievable rate and mitigates the near-target blind zone issue. Moreover, by applying a similar design philosophy, the authors of \cite{TVT_2022_Zhang_Integrated} developed a scheme that embeds communication information into the support of one sparse vector and transmits a low-dimensional signal via a spreading codebook. In addition, to realize an adaptive frame structure configuration for sensing and communication dual functions, the authors of \cite{JSTSP_2021_Zhang_Design} elaborately designed an ISAC system based on the 5G New Radio protocol. Besides, several research progresses have sprung up to combine ISAC with other emerging technologies, such as non-orthogonal multiple access \cite{CL_2022_Wang_NOMA}, deep learning \cite{CL_2021_Mu_Integrated}, orthogonal time frequency space modulation \cite{JSTSP_2021_Yuan_Integrated}, holographic MIMO \cite{JSAC_2022_Zhang_Holographic, ICC_2023_An_Stacked}, and reconfigurable intelligent surface \cite{TVT_2022_Wang_Joint, TCOM_2022_An_Low}, etc., to unlock the full potential of ISAC.

Despite these interesting ISAC research efforts, many key issues about ISAC remain unexplored, such as a unified analytical framework, theoretical performance bounds, optimal signal processing algorithms, etc. \cite{JSTSP_2021_An_Zhang}. In particular, characterizing the fundamental performance limits and the inherent tradeoffs between sensing and communication functionalities is of great importance for the future development of ISAC technologies \cite{CST_2022_Liu_A}. Specifically, investigating the fundamental performance limits can doubtlessly reveal the potential gaps between the current ISAC technologies and the optimal designs. Moreover, portraying performance tradeoffs is capable of providing useful guidance and insights for protocol design and theoretical analysis of practical ISAC schemes. At the time of writing, a few works have been dedicated to analyzing the theoretical performance of ISAC systems. Specifically, a systematic classification method for ISAC technologies was proposed in \cite{CST_2022_Liu_A}, based on which unified order-wise expressions for the Cr\'amer-Rao lower bound (CRLB) of sensing parameters were provided. Furthermore, the authors of \cite{JSAC_2022_Liu_Integrated} discussed a couple of performance tradeoffs in ISAC systems, e.g., the successful probability of detection (PD) versus (vs.) communication rate, and radar estimation rate \cite{TSP_2016_Chiriyath_Inner} vs. channel capacity. Besides, in \cite{JSAC_2022_Xiao_Waveform}, the authors analyzed the PD and the ambiguity function for sensing, as well as the spectral efficiency of an FD-ISAC scheme by taking into account the residual self-interference. More recently, the diversity orders of a general ISAC system were analyzed in \cite{arXiv_2022_Ouyang_Fundamental} for both the communication rate and sensing rate. As for the ISAC scenario involving a moving target, \cite{arXiv_2022_Liu_Predictive} derived the CRLBs of motion parameter estimation to quantify the sensing performance, based on which a sum-rate maximization problem with CRLB-based sensing constraints was formulated. To a step further, the PD, the CRLB for parameter estimation as well as the posterior CRLB for moving target indication were derived in \cite{arXiv_2022_Dong_Sensing} to measure the sensing QoS for detection, localization, and tracking, respectively. Motivated readers may refer to \cite{CST_2022_Liu_A, JSAC_2022_Liu_Integrated, arXiv_2021_Ahmadipour_An, JSAC_2022_Shi_Device} and references therein for more details on performance analysis of ISAC systems.

\begin{table*}[!t]\label{tab1}
\centering
\scriptsize
\caption{Comparison between our contributions and existing work on ISAC performance analysis.}
\begin{threeparttable}
\begin{tabular}{l||c||c c c c c c c c c c}
\hline 
 Scenarios & Our work & \cite{CST_2022_Liu_A} & \cite{JSAC_2022_Liu_Integrated} & \cite{JSAC_2022_Xiao_Waveform} & \cite{arXiv_2022_Ouyang_Fundamental} & \cite{arXiv_2022_Liu_Predictive} & \cite{arXiv_2022_Dong_Sensing} & \cite{arXiv_2021_Ahmadipour_An} & \cite{JSAC_2022_Shi_Device} \\ \hline\hline
 Sensing-free communication & \CheckmarkBold & \XSolidBrush & \CheckmarkBold & \CheckmarkBold & \CheckmarkBold & \CheckmarkBold & \CheckmarkBold & \CheckmarkBold & \XSolidBrush \\ 
 Sensing-interfered communication & \CheckmarkBold & \CheckmarkBold & \XSolidBrush & \XSolidBrush & \XSolidBrush & \XSolidBrush & \XSolidBrush & \CheckmarkBold & \XSolidBrush \\ 
 Communication-assisted sensing & \CheckmarkBold & \XSolidBrush & \XSolidBrush & \XSolidBrush & \XSolidBrush & \XSolidBrush & \XSolidBrush & \XSolidBrush & \CheckmarkBold \\ 
 Communication-interfered sensing & \CheckmarkBold & \CheckmarkBold & \XSolidBrush & \CheckmarkBold & \CheckmarkBold & \CheckmarkBold & \CheckmarkBold & \CheckmarkBold & \XSolidBrush \\ \hline \hline
{Closed-form expressions for PFA \& PD} & \CheckmarkBold & \XSolidBrush & \CheckmarkBold & \CheckmarkBold & \XSolidBrush & \XSolidBrush & \XSolidBrush & \XSolidBrush & \XSolidBrush \\ 
{Power allocation} & \CheckmarkBold & \XSolidBrush & \XSolidBrush & \XSolidBrush & \XSolidBrush & \XSolidBrush & \CheckmarkBold & \XSolidBrush & \XSolidBrush \\ 
{Performance tradeoff analysis} & \CheckmarkBold & \CheckmarkBold & \CheckmarkBold & \CheckmarkBold & \CheckmarkBold & \XSolidBrush & \CheckmarkBold & \CheckmarkBold & \XSolidBrush \\ \hline\hline
{Sensing performance metric} & PD & CRLB & PD & PD & Sensing rate & CRLB & CRLB & Distortion & \multicolumn{2}{c}{\makecell[c]{Location\\accuracy}} \\ \hline
{Communication performance metric} & Rate & Capacity & Rate & Spectral efficiency & Ergodic rate & Sum-rate & Sum-rate & Capacity & None \\ \hline
{ISAC system setup} & SISO\tnote{1} & MIMO & MIMO & Full-duplex & MIMO & V2I\tnote{2} & MIMO & SISO & OFDM\tnote{3} \\ \hline
\end{tabular}
\begin{tablenotes}
\item[1] SISO: Single-input single-output \item[2] V2I: Vehicle-to-infrastructure \item[3] OFDM: Orthogonal frequency division multiplexing
\end{tablenotes}
\dotfill
\end{threeparttable}
\end{table*}

Nevertheless, most of the existing analytical results considered only the \emph{ad-hoc} ISAC systems to formulate and solve the corresponding waveform design and resource allocation problems therein, thus lacking a general framework to quantitatively characterize the interrelation between sensing and communication functionalities. For instance, the performance tradeoff analysis in \cite{JSAC_2022_Liu_Integrated} employed the PD derived for the radar and communication (R\&C) coexistence system \cite{SPL_2017_Chalise_Performance}, in which the sensing and communication waveforms are allocated orthogonal resource blocks. Additionally, the theoretical analysis presented in \cite{JSAC_2022_Xiao_Waveform} is limited to the FD-ISAC design, which lacks a general discussion on the performance tradeoff analysis. In this paper, we aim to fill this gap by presenting a comprehensive framework that allows for quantitatively analyzing the fundamental tradeoff between the PD for sensing and the achievable rate of a communication user (CU). To the best of our knowledge, this is the first attempt to gain a well-rounded insight by investigating the intrinsic performance tradeoff between sensing and communication functionalities in an ISAC system. For the sake of illustration, we boldly and explicitly contrast the contributions of this paper to other works on ISAC performance analysis in Table \ref{tab1}. Specifically, our main contributions are further summarized as follows:

\begin{itemize}
 \item Firstly, we present a generalized framework for analyzing the performance of an ISAC system consisting of one ISAC base station (BS) serving a single CU and simultaneously monitoring a target of interest. Following this, we define four typical scenarios in the ISAC system considered: \emph{i)} sensing-free communication; \emph{ii)} sensing-interfered communication; \emph{iii)} communication-assisted sensing; and \emph{iv)} communication-interfered sensing.
 \item Secondly, we analyze the achievable rates of the CU under the sensing-free and sensing-interfered communication scenarios. With regard to the sensing performance, we derive closed-form expressions for both the probability of false alarm (PFA) and the PD considering communication-assisted and communication-interfered sensing scenarios. In our theoretical analysis, the effects of the unknown channel coefficient are also taken into account. Note that in sharp contrast to the conventional R\&C coexistence system where sensing and communication waveforms are orthogonal, our communication-interfered sensing scenario considers the interference caused by the communication waveforms, rendering it challenging to obtain the closed-form expressions of PD and PFA. To solve this issue, we utilize the Chebyshev-Gaussian quadrature to numerically obtain an approximated expression instead.
 \item Thirdly, we formulate a power allocation problem to characterize the fundamental tradeoff between the PD and the achievable rate, where the objective is to minimize the transmit power at the ISAC-BS, while ensuring the PD required for monitoring the target of interest and the CU's requirement on achievable rate. We elaborate on eight typical ISAC scenarios by combining the aforementioned two communication modes and two sensing modes. The optimal power allocation solution under each ISAC scenario is provided. Since in some cases it is challenging to obtain an explicit power allocation solution due to the transcendental equations, we provide parametric expression for the optimal solution instead, where the underlying parameter is numerically obtained.
 \item Finally, extensive simulation results corroborate our theoretical analysis and verify the effectiveness of the proposed power allocation solutions. It is demonstrated that in a collaborative ISAC system, the sensing and communication capabilities could achieve mutual gain from each other, whereas there exists an intrinsic tradeoff when they operate in a competitive manner. Moreover, our simulation results also reveal the performance advantages of an ISAC system over a conventional R\&C coexistence counterpart.

\end{itemize}

The rest of this paper is organized as follows. In Section \ref{sec2}, we introduce the general ISAC system model. Then, Section \ref{sec3} evaluates the communication performance and sensing performance, in which the PFA and the PD under various sensing scenarios are derived. Furthermore, the power allocation problems are investigated in Section \ref{sec4}. Finally, Section \ref{sec5} provides numerical results to verify our analysis before concluding the paper in Section \ref{sec6}.

\emph{Notations:} We use italic, bold lowercase, and bold uppercase letters to denote scalars, vectors, and matrices, respectively. ${\left( \cdot \right)^*}$, ${\left( \cdot \right)^T}$, and ${\left( \cdot \right)^H}$ represent the conjugate, transpose, and Hermitian transpose, respectively. $\left| z \right|$, $\angle z$, $\Re\left( z \right)$, and $\Im\left( z \right)$ refer to the magnitude, phase, real part, and imaginary part, respectively, of a complex number $z$. $\left\| \mathbf{v} \right\|$ is the ${l_2}$-norm of a complex vector $\mathbf{v}$. $\mathbb{E}\left( \cdot \right)$ stands for the expectation operation. $\log_{a} \left( \cdot \right)$ is the logarithmic function with base $a$, while $\ln \left( \cdot \right)$ is the natural logarithm. ${{\mathbb{C}}^{x \times y}}$ represents the space of $x \times y$ complex-valued matrices. Furthermore, ${\bf{0}}$ and ${\bf{1}}$ denote all-zero and all-one vectors, respectively, with appropriate dimensions, while ${{\bf{I}}_N}$ denotes the $N \times N$ identity matrix. $a!$ is the factorial of a non-negative integer $a$. The distribution of a circularly symmetric complex Gaussian (CSCG) random vector with mean vector ${\boldsymbol{\mu }}$ and covariance matrix ${\boldsymbol{\Sigma }}$ is denoted by $ \sim {\mathcal{CN}}\left( {{\boldsymbol{\mu }},{\boldsymbol{\Sigma }}} \right)$, where $ \sim$ stands for ``distributed as''. The distribution of a real-valued Gaussian random vector with mean ${\mu }$ and variance ${\sigma ^2}$ is denoted by $ \sim {\mathcal{N}}\left( {{\mu },{\sigma ^2}} \right)$, while the non-central Chi-square distribution with $k$ degrees-of-freedom and non-centrality parameter $\beta$ is represented by $\sim \chi _k^2\left( \beta \right)$. Moreover, $\gamma \left ( s,x \right )=\int_{0}^{x}t^{s-1}e^{-t}dt$ and $\Gamma \left ( z \right )=\int_{0}^{\infty }x^{z-1}e^{-x}dx$ denote the lower incomplete and original gamma function, respectively. $Q\left( x \right) = \frac{1}{{\sqrt {2\pi } }}\int_x^\infty {{e^{ - \frac{{{t^2}}}{2}}}dt}$ is the $\text{Q}$-function. ${Q_m}\left( {a,b} \right) = \frac{1}{{{a^{m - 1}}}}\int_b^\infty {{x^m}{e^{ - \frac{{{x^2} + {a^2}}}{2}}}{I_{m - 1}}\left( {ax} \right)dx} $ is the generalized Marcum $\text{Q}$-function of order $m$ for non-centrality parameter $a$, in which ${I_m}\left( \cdot \right)$ denotes the modified Bessel function of the first kind of order $m$.
\section{System Model}\label{sec2}
\subsection{Transmit Signal Model at the ISAC-BS}
\begin{figure*}[!t]
\centering
\includegraphics[width=16cm]{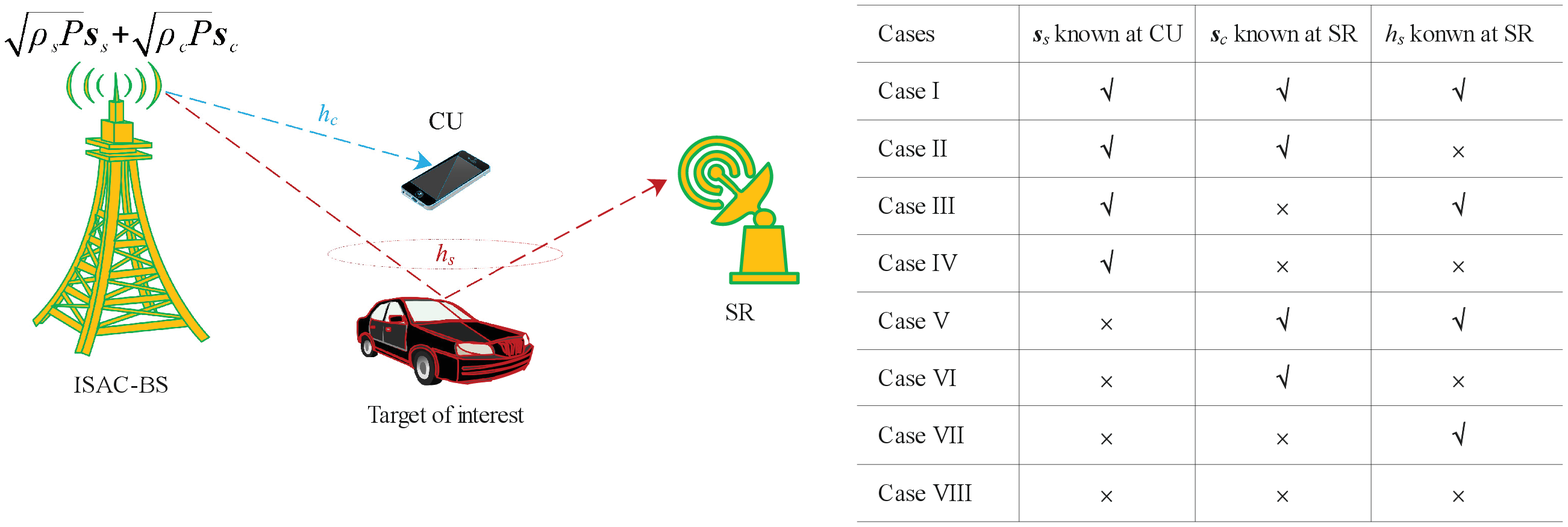}
\caption{A downlink ISAC system with an ISAC-BS serving a single CU and monitoring a target of interest, where we consider eight typical ISAC cases.}
\label{fig1}
\end{figure*}
As shown in Fig. \ref{fig1}, we consider a downlink ISAC system, where an ISAC-BS is deployed to support both communications towards a single CU and the task of sensing a target of interest simultaneously\footnote{Here we are analyzing an elementary ISAC scenario consisting a single user and a single target for unveiling insights into the performance tradeoff between sensing and communication functionalities. The specific details in more complex ISAC scenarios involving multiple users and targets (e.g., beamforming, power allocation, interference mitigation, performance tradeoff analysis, etc.) will be the subject of our future work.}. Additionally, a sensing receiver (SR) is deployed to collect echoes reflected by the target for sensing its state\footnote{Although the bistatic sensing mode with a separated BS and SR deployment is adopted, our subsequent theoretical analysis is certainly applicable for the ISAC scenario with a co-site BS and SR deployment.}, e.g., determining the presence/absence of the target, acquiring the vehicle attitude information, etc. In this paper, we focus on the detection task of the ISAC system, where the SR determines the presence/absence of the target of interest by matching the echoes with the sensing signal locally regenerated. For the sake of elaboration, we assume that both the BS, the CU, and the SR are equipped with a single antenna. Moreover, we consider a clutter-free environment by assuming that the clutter mitigation has been addressed through a variety of existing techniques, e.g., \cite{TGRS_2016_Xu_Space}.

In the downlink transmission, the ISAC-BS allocates a portion of the total power to broadcast a sensing-aimed waveform, ${s_s}\left( t \right)\in \mathbb{C}$ at the $t$-th slot, while the remaining power is employed for transmitting an information-bearing signal, ${s_c}\left( t \right)\in \mathbb{C}$ at the $t$-th slot, to the CU. Here, we assume that both ${s_s}\left( t \right)$ and ${s_c}\left( t \right)$ are normalized, i.e., ${\mathbb{E}}\left( {{{\left| {{s_s}\left( t \right)} \right|}^2}} \right) = {\mathbb{E}}\left( {{{\left| {{s_c}\left( t \right)} \right|}^2}} \right) = 1$, and statistically independent of each other, without loss of generality. In contrast to most conventional R\&C coexistence systems that employ orthogonal resource blocks (time, frequency, beam, etc.) to schedule these two signal transmissions \cite{SPM_2019_Zheng_Radar}, we consider a general ISAC design where the sensing and communication waveforms are superimposed over the same time-frequency resources \cite{WCL_2022_Xu_Deep, TSP_2019_Wang_Power}. Accordingly, the downlink normalized signal transmitted from the ISAC-BS at the $t$-th slot is given by
\begin{align}
 s\left( t \right) = \sqrt {{\rho _s}} {s_s}\left( t \right) + \sqrt {{\rho _c}} {s_c}\left( t \right),
\end{align}
where ${{\rho _s}} $ and ${{\rho _c}} $ are the non-negative normalized power coefficients allocated for sensing the target and communicating with the CU, respectively, satisfying ${\rho _s} + {\rho _c} = 1$.

In this paper, we consider the narrowband quasi-static block fading channel for the BS-CU link \cite{Book_2005_Tse_Fundamentals}. Let $h_c\in \mathbb{C}$ denote the channel from the ISAC-BS to the CU and the signal received at the CU is thus expressed as\footnote{Note that in \eqref{eq2} we have implicitly assumed that the signal reflected by the target to the CU is sufficiently weak compared to that coming through the direct BS-CU path and thus is negligible. For the case where the target plays a scattering role in the communication link from the ISAC-BS to the CU, please refer to \cite{CST_2022_Liu_A} for some basic insights.}
\begin{align}
{r_c}\left( t \right) &= {h_c}\sqrt P s\left( t \right) + {n_c}\left( t \right) \notag \\
&= \sqrt {{\rho _s}P} {h_c}{s_s}\left( t \right) + \sqrt {{\rho _c}P} {h_c}{s_c}\left( t \right) + {n_c}\left( t \right), \label{eq2}
\end{align}
where $P$ denotes the total power available at the ISAC-BS, while ${n_c}\left( t \right)\sim {\mathcal{CN}}\left( {0,\sigma _c^2} \right)$ is the additive white Gaussian noise (AWGN), with $\sigma _c^2$ denoting the noise power at the CU's receiver.

Furthermore, let $h_s\in \mathbb{C}$ denote the composite BS-target-SR channel coefficient, and the signal received at the SR is given by
\begin{align}
{r_s}\left( t \right) =& {h_s}\sqrt P s\left( {t - \tau } \right){e^{j2\pi {f_d}t}} + {n_s}\left( t \right) \notag \\
=& \sqrt {{\rho _s}P} {h_s}{s_s}\left( {t - \tau } \right){e^{j2\pi {f_d}t}} \notag \\
&+ \sqrt {{\rho _c}P} {h_s}{s_c}\left( {t - \tau } \right){e^{j2\pi {f_d}t}} + {n_s}\left( t \right), \label{eq3}
\end{align}
where $\tau$ and ${f_d}$ denote the delay and Doppler shift induced by the motion of the target, respectively, ${n_s}\left( t \right)\sim {\mathcal{CN}}\left( {0,\sigma _s^2} \right)$ is the AWGN at the SR with the receiver sensitivity of $\sigma _s^2$.

By examining \eqref{eq2} and \eqref{eq3}, one may note that in addition to the power competition between sensing and communication functionalities as in the conventional R\&C coexistence systems, there also exists the waveform interaction that impacts each other's performance, resulting in a fundamental tradeoff between the sensing and communication performance since they share the same radio resources and hardware equipment. In Section \ref{sec4}, we will consider eight typical ISAC scenarios to investigate the underlying performance tradeoffs, as listed in Fig. \ref{fig1}.

\subsection{Receive Signal Processing at the SR}
Specifically, we focus on the target detection problem for acquiring fundamental insights into the intrinsic performance tradeoff between the sensing and communication functionalities. In particular, PD is employed as a figure of merit for sensing in our theoretical analysis framework.

For the ISAC system involving a single target, the detection problem can be formulated as a binary hypothesis testing problem \cite{SPL_2017_Chalise_Performance}. Under the null hypothesis, ${\mathcal{H}_0}$, and the alternative hypothesis, ${\mathcal{H}_1}$, the received signals of $T$ samples at the SR can be expressed as
\begin{align}
\left\{ {\begin{array}{*{20}{l}}
{{{\bf{r}}_s} = {{\bf{n}}_s},}&{{\mathcal{H}_0}},\\
{{{\bf{r}}_s} = {h_s}\sqrt P {\bf{Ds}} + {{\bf{n}}_s},}&{{\mathcal{H}_1}},
\end{array}} \right.
\end{align}
respectively, where ${\bf{D}} \in {{\mathbb{C}}^{T \times T}}$ is a unitary delay-Doppler operator matrix determined by $\tau$ and ${f_d}$, while ${\bf{s}} \in {{\mathbb{C}}^{T \times 1}}$ is a vector collecting sampled ISAC waveform transmitted from the BS, ${{{\bf{n}}_s}} \sim {\mathcal{CN}}\left( {{{\bf{0}}_{T \times 1}},\sigma _s^2{{\bf{I}}_T}} \right)$ is the noise vector at the SR. For known locations of the stationary transmitter and receiver, as well as the location and velocity of the target at a hypothesized position in the delay-Doppler cell, ${\bf{D}}$ can be readily calculated \cite{SPL_2017_Chalise_Performance}. Moreover, since ${{\bf{D}}^H}{\bf{D}} = {\bf{I}}$, the received signals after unitary transformation with ${\bf{D}}$ become
\begin{align}
\left\{ {\begin{array}{*{20}{l}}
{{{{\bf{\tilde r}}}_s} = {{{\bf{\tilde n}}}_s},}&{{\mathcal{H}_0}},\\
{{{{\bf{\tilde r}}}_s} = {h_s}\sqrt P {\bf{s}} + {{{\bf{\tilde n}}}_s},}&{{\mathcal{H}_1}},
\end{array}} \right.
\end{align}
where we have ${{{\bf{\tilde r}}}_s} = {{\bf{D}}^H}{{\bf{r}}_s}$ and ${{{\bf{\tilde n}}}_s} = {{\bf{D}}^H}{{\bf{n}}_s} \sim {\mathcal{CN}}\left( {{{\bf{0}}_{T \times 1}},\sigma _s^2{{\bf{I}}_T}} \right)$.

Furthermore, the target detection is performed by comparing the likelihood ratio test (LRT) function defined by
\begin{align}
\Lambda \left( {{{{\bf{\tilde r}}}_s}} \right) = \frac{{f\left( {{{{\bf{\tilde r}}}_s}\left| {{\mathcal{H}_1}} \right.} \right)}}{{f\left( {{{{\bf{\tilde r}}}_s}\left| {{\mathcal{H}_0}} \right.} \right)}}, \label{eq9}
\end{align}
with a certain threshold value $\overline{\kappa }$, which is determined according to a target value of PFA for performing constant false alarm rate (CFAR) detection \cite{Book_2013_Poor_An}, ${f\left( {{{{\bf{\tilde r}}}_s}\left| {{\mathcal{H}_1}} \right.} \right)}$ and ${f\left( {{{{\bf{\tilde r}}}_s}\left| {{\mathcal{H}_0}} \right.} \right)}$ denote the probability density functions (PDFs) of ${{{{\bf{\tilde r}}}_s}}$ under the hypotheses ${\mathcal{H}_1}$ and ${\mathcal{H}_0}$, respectively, which will be defined in Section \ref{sec3_2} when considering specific sensing scenarios.

\emph{Remark 1:} Before proceeding further, we elaborate on the practical target detection process. In general, the target detection is carried out simultaneously with the estimation of delay and Doppler shift \cite{TSP_2014_Liu_Two, SJ_2017_Zaimbashi_Target}. By estimating and compensating for the delay and Doppler shift via some grid-based searching approaches as used in conventional active radars \cite{TSP_2014_Liu_Two}, one could attain the optimal coherent detection performance. Again, we highlight that this paper focuses on the evaluation of the fundamental tradeoff between the target detection probability and the achievable rate. The imperfect estimates of delay and Doppler shift would confuse our theoretical analysis. For this reason, we assume that both the delay and Doppler shift have been perfectly compensated \cite{SJ_2017_Zaimbashi_Target}. Motivated readers can refer to \cite{SJ_2017_Zaimbashi_Target, TSP_2013_Palmer_DVB} for gaining more about the effects of delay and Doppler shift on coherent detection. Nonetheless, it is worth noting that our communication-assisted sensing with the use of the estimated communication waveform turns out to be the classic energy detector, i.e., \eqref{eq22}, which corresponds to the case without compensating for the delay and Doppler shift.

\section{Performance Evaluation of an ISAC system}\label{sec3}
In this section, we evaluate the sensing and communication performance in an ISAC system. Specifically, we focus on two distinct communication scenarios (i.e., sensing-free and sensing-interfered), as well as two types of sensing scenarios (i.e., communication-assisted and communication-interfered).

\subsection{Communication Performance Evaluation}
We commence by evaluating the communication performance in terms of the achievable rate at the CU. Specifically, we consider the sensing-free and sensing-interfered communication scenarios, in which the sensing signal ${s_s}\left( t \right)$ is regarded as \emph{a priori} known and unknown, respectively, by the CU.
\subsubsection{Sensing-Free Communication}\label{sec3_1}
For the CU legitimately admitted into the RAN, it is reasonable to assume that the sensing signal ${s_s}\left( t \right)$ specified by the protocol is \emph{a priori} known by the CU. Accordingly, the portion of the sensing signal can be removed at the receiver by employing the successive interference cancellation (SIC) technique. We suppose that the sensing signal is completely mitigated, provided that the perfect channel state information (CSI) is available at the receiver. Thus, we have
\begin{align}
    {s_c}\left( t \right) = \frac{h_{c}^{*}}{\sqrt{{\rho _c}P}\left | h_{c} \right |^{2}}{r_c}\left( t \right) - \sqrt {\frac{{\rho _s}}{{\rho _c}}} {s_s}\left( t \right).
\end{align}

As a result, the instantaneous achievable rate in terms of bits-per-second-per-Hertz (b/s/Hz) at the CU can be expressed by \cite{TVT_2022_An_The}
\begin{align}
 R = {\log _2}\left( {1 + \frac{{{\rho _c}P{{\left| {{h_c}} \right|}^2}}}{{\sigma _c^2}}} \right). \label{eq4}
\end{align}

\subsubsection{Sensing-Interfered Communication}
For the CU having a conventional receiver that lacks the capability to perform SIC, the CU suffers from extra interference caused by the sensing signal transmitted from the ISAC-BS. Therefore, the instantaneous achievable rate at the CU is given by \cite{GLOBECOM_2021_Hua_Transmit}
\begin{align}
R = {\log _2}\left( {1 + \frac{{{\rho _c}P{{\left| {{h_c}} \right|}^2}}}{{{\rho _s}P{{\left| {{h_c}} \right|}^2} + \sigma _c^2}}} \right). \label{eq5}
\end{align}
\subsection{Sensing Performance Evaluation}\label{sec3_2}
Next, we proceed to evaluate the sensing performance in our ISAC framework by taking into account the effects of the communication signal. Throughout this paper, we consider the device-based sensing scenario where the sensing signal is always known at the SR.
\subsubsection{Communication-Assisted Sensing}\label{sec3-2-1}
For a given modulation type specified by the communication protocol, to improve the performance of the target detection, it is straightforward to recover the communication waveform first from the signal received at the SR. Specifically, the SIC technique is invoked for mitigating the interference caused by the sensing signal before carrying out the demodulation. Let ${{{\bf{s}}_s}} \in {{\mathbb{C}}^{T \times 1}}$ and ${{{\bf{s}}_c}} \in {{\mathbb{C}}^{T \times 1}}$ denote the sensing signal vector and communication signal vector, respectively. The recovered communication signal can thus be expressed as
\begin{align}
 {\hat {\mathbf{s}}_c} = \mathbb{Q}\left( {\frac{h_{s}^{*}}{{\sqrt {{\rho _c}P}\left | h_{s} \right |^{2} }}{\tilde {\mathbf{r}}_s} - \sqrt {\frac{{{\rho _s}}}{{{\rho _c}}}} {{\mathbf{s}}_s}} \right), \label{eq8}
\end{align}
where $\mathbb{Q}:\mathbb{C}^{T}\rightarrow \mathbb{S}^{T}$ denotes the slicing operator for demodulation with $\mathbb{S}$ representing the constellation of the communication symbol. By doing so, the communication waveform can be exploited to facilitate target detection.

For the sake of elaboration, let us first consider the case where the communication signal ${{\mathbf{s}}_c}$ is perfectly recovered by \eqref{eq8} to obtain an upper bound of the detection performance. In this case, we have
\begin{align}
f\left( {{{{\bf{\tilde r}}}_s}\left| {{\mathcal{H}_1}} \right.} \right) &= \frac{1}{{{{\left( {\pi \sigma _s^2} \right)}^T}}}{{\rm{e}}^{ - \frac{{{{\left\| {{{{\bf{\tilde r}}}_s} - {h_s}\sqrt P {\bf{s}}} \right\|}^2}}}{{\sigma _s^2}}}}. \label{eq10}\\
f\left( {{{{\bf{\tilde r}}}_s}\left| {{\mathcal{H}_0}} \right.} \right) &= \frac{1}{{{{\left( {\pi \sigma _s^2} \right)}^T}}}{{\mathop{\rm e}\nolimits} ^{ - \frac{{{{\left\| {{{{\bf{\tilde r}}}_s}} \right\|}^2}}}{{\sigma _s^2}}}}. \label{eq11}
\end{align}
Note that for the communication-assisted sensing mode, the superimposed sensing and communication signal received at the SR is exploited to perform the target detection after recovering the communication waveform ${\mathbf{s}}_c$, which is similar to the classic passive radar system that utilizes the available reference channel \cite{SPL_2017_Chalise_Performance}.

Next, we will examine two possible scenarios that depend on whether $h_s$ is known at the SR. Furthermore, we will provide a shared lower bound for these two scenarios by utilizing an estimated ${\mathbf{s}}_c$.
\paragraph{${h_s}$ known at the SR}
Given the case where ${h_s}$ is known at the SR, by substituting \eqref{eq10} $\sim$ \eqref{eq11} into \eqref{eq9} and taking its logarithm, the logarithmic LRT function becomes
\begin{align}
\ln\Lambda \left( {{{{\bf{\tilde r}}}_s}} \right) &= \frac{1}{{\sigma _s^2}}\left( {{{\left\| {{{{\bf{\tilde r}}}_s}} \right\|}^2} - {{\left\| {{{{\bf{\tilde r}}}_s} - {h_s}\sqrt P {\bf{s}}} \right\|}^2}} \right) \notag \\
&=\frac{1}{{\sigma _s^2}}\left( {2\sqrt P \Re \left( {{h_s}{\bf{\tilde r}}_s^H{\bf{s}}} \right) - P{{\left\| {{h_s}{\bf{s}}} \right\|}^2}} \right). \label{eq12}
\end{align}

Based on \emph{Eq. (II.B.26)} of \cite{Book_2013_Poor_An}, ${\Re\left( {{h_s}{\bf{\tilde r}}_s^H{\bf{s}}} \right)}$ in \eqref{eq12} under hypotheses ${\mathcal{H}_0}$ and ${\mathcal{H}_1}$, respectively, are distributed as
\begin{align}
\left\{ {\begin{array}{*{20}{l}}
{\Re \left( {{h_s}{\bf{\tilde r}}_s^H{\bf{s}}} \right)\sim {\cal N}\left( {0,\frac{1}{2}\sigma _s^2{{\left\| {{h_s}{\bf{s}}} \right\|}^2}} \right),}&{{\mathcal{H}_0}},\\
{\Re \left( {{h_s}{\bf{\tilde r}}_s^H{\bf{s}}} \right)\sim {\cal N}\left( {\sqrt P {{\left\| {{h_s}{\bf{s}}} \right\|}^2},\frac{1}{2}\sigma _s^2{{\left\| {{h_s}{\bf{s}}} \right\|}^2}} \right),}&{{\mathcal{H}_1}}.
\end{array}} \right.
\end{align}

Therefore, the PFA can be readily calculated by
\begin{align}\label{eq14}
{P_{\textrm{FA}}} &= \text{Pr}\left\{ {\left. {\frac{1}{{\sigma _s^2}}\left( {2\sqrt P \Re \left( {{h_s}{\bf{\tilde r}}_s^H{\bf{s}}} \right) - P{{\left\| {{h_s}{\bf{s}}} \right\|}^2}} \right) \ge \kappa } \right|{\mathcal{H}_0}} \right\} \notag \\
&= Q\left( {\frac{{\sigma _s^2\kappa + P{{\left\| {{h_s}{\bf{s}}} \right\|}^2}}}{{\sqrt {2P} {\sigma _s}\left\| {{h_s}{\bf{s}}} \right\|}}} \right) \mathop \simeq \limits^{T \gg 1} Q\left( {\frac{{\sigma _s^2\kappa + PT{{\left| {{h_s}} \right|}^2}}}{{\sqrt {2PT} {\sigma _s}\left| {{h_s}} \right|}}} \right),
\end{align}
where we have $\kappa = \ln\left( \bar \kappa \right)$. Note that the approximation is valid because we have $\frac{1}{T}\left \| \mathbf{s} \right \|^{2}=1$ for $T\gg 1$.

Similarly, the PD is expressed as
\begin{align}
{P_{\textrm{D}}} &= \text{Pr}\left\{ {\left. {\frac{1}{{\sigma _s^2}}\left( {2\sqrt P \Re \left( {{h_s}{\bf{\tilde r}}_s^H{\bf{s}}} \right) - P{{\left\| {{h_s}{\bf{s}}} \right\|}^2}} \right) \ge \kappa } \right|{\mathcal{H}_1}} \right\} \notag \\
&= Q\left( {\frac{{\sigma _s^2\kappa - P{{\left\| {{h_s}{\bf{s}}} \right\|}^2}}}{{\sqrt {2P} {\sigma _s}\left\| {{h_s}{\bf{s}}} \right\|}}} \right)\mathop \simeq \limits^{T \gg 1} Q\left( {\frac{{\sigma _s^2\kappa - PT{{\left| {{h_s}} \right|}^2}}}{{\sqrt {2PT} {\sigma _s}\left| {{h_s}} \right|}}} \right). \label{eq15}
\end{align}

Note that \eqref{eq14} and \eqref{eq15} have similar expressions to that in the classic coherent detector \cite{Book_2001_Simon_Digital}, which we included here for maintaining the content integrity.
\paragraph{${h_s}$ unknown at the SR}
Next, let us consider a more practical ISAC scenario where $h_s$ is unknown at the SR\footnote{Note that when considering the case where the SR is integrated with the BS, the bi-static sensing mode is transformed into a monostatic one, and the communication waveform, ${{\mathbf{s}}_c}$, becomes fully known by the BS (SR) without the need for performing the estimation procedure as in \eqref{eq8}. Nevertheless, one might still face the case with an unknown channel coefficient $h_s$. Here, we consider the case with unknown ${h_s}$ in order to provide a comprehensive framework for all potential ISAC scenarios.}. In this case, $h_s$ is substituted with its estimated value in the LRT function, which leads to a new test function known as the generalized LRT (GLRT) \cite{SPL_2017_Chalise_Performance}. Specifically, taking the logarithm of \eqref{eq10}, the maximum likelihood estimate of ${h_s}$ can be obtained by
\begin{align}
{\hat{h}_s} &= \arg \mathop {\max }\limits_{{h_s}} \ln f\left( {{{{\bf{\tilde r}}}_s}\left| {{\mathcal{H}_1}} \right.} \right) \notag \\
&= \arg \mathop {\min }\limits_{{h_s}} {\left\| {{{{\bf{\tilde r}}}_s} - {h_s}\sqrt P {\bf{s}}} \right\|^2} = \frac{1}{{\sqrt P {{\left\| {\bf{s}} \right\|}^2}}}{{\bf{s}}^H}{{{\bf{\tilde r}}}_s}. \label{16}
\end{align}

Substituting \eqref{16} into \eqref{eq12} yields the logarithmic GLRT function as
\begin{align}
\ln\Lambda \left( {{{{\bf{\tilde r}}}_s}} \right) &= \frac{1}{{\sigma _s^2}}\left( {{{\left\| {{{{\bf{\tilde r}}}_s}} \right\|}^2} - {{\left\| {{{{\bf{\tilde r}}}_s} - \frac{{{\bf{s}}{{\bf{s}}^H}}}{{{{\left\| {\bf{s}} \right\|}^2}}}{{{\bf{\tilde r}}}_s}} \right\|}^2}} \right) \notag \\
&= \frac{1}{{\sigma _s^2}}{{{\bf{\tilde r}}}_s}^H{\bf{G}}{{{\bf{\tilde r}}}_s} = \frac{1}{{\sigma _s^2}}{{{\bf{\tilde r}}}_s}^H{\bf{U\Lambda }}{{\bf{U}}^H}{{{\bf{\tilde r}}}_s},\label{eq17}
\end{align}
where we have ${\bf{G}} = {\bf{I}} - {\left( {{\bf{I}} - \frac{{{\bf{s}}{{\bf{s}}^H}}}{{{{\left\| {\bf{s}} \right\|}^2}}}} \right)^H}\left( {{\bf{I}} - \frac{{{\bf{s}}{{\bf{s}}^H}}}{{{{\left\| {\bf{s}} \right\|}^2}}}} \right)$, ${\bf{\Lambda }}$ is a diagonal matrix consisting of the eigenvalues of ${\bf{G}}$, while ${\bf{U}} \in \mathbb{C}^{T\times T}$ is a unitary matrix whose columns are the corresponding eigenvectors. It is evident that only one of the eigenvalue in ${\bf{\Lambda }}$ is $1$, while the remaining $\left( {T - 1} \right)$ eigenvalues are $0$.

Based on \eqref{eq17}, the exact closed-form expression for PFA is given by
\begin{align}\label{eq18}
{P_{\textrm{FA}}} &= \text{Pr}\left\{ {\left. {\frac{1}{{\sigma _s^2}}{{{\bf{\tilde r}}}_s}^H{\bf{U\Lambda }}{{\bf{U}}^H}{{{\bf{\tilde r}}}_s} \ge \kappa } \right|{\mathcal{H}_0}} \right\} \notag \\
&= \text{Pr}\left\{ {{\bf{n'}}_s^H{\bf{\Lambda }}{{{\bf{n'}}}_s} \ge \sigma _s^2\kappa } \right\} \notag \\
&= \text{Pr}\left\{ {{{\left| {{{n}_{s,1}'}} \right|}^2} \ge \sigma _s^2\kappa } \right\} = {e^{ - \kappa }},
\end{align}where we have ${{\bf{n'}}_s} = {{\bf{U}}^H}{{\bf{\tilde n}}_s}\sim {\cal C}{\cal N}\left( {{{\bf{0}}_{T \times 1}},\sigma _s^2{{\bf{I}}_T}} \right)$.

Furthermore, the PD in this case can be expressed as
\begin{align}
{P_{\textrm{D}}} &= \text{Pr}\left\{ {\left. {\frac{1}{{\sigma _s^2}}{\bf{\tilde r}}_s^H{\bf{U\Lambda }}{{\bf{U}}^H}{{{\bf{\tilde r}}}_s} \ge \kappa } \right|{\mathcal{H}_1}} \right\} \notag \\
&= \text{Pr}\left\{ {{{\left( {{h_s}\sqrt P {\bf{s'}} + {{{\bf{n'}}}_s}} \right)}^H}{\bf{\Lambda }}\left( {{h_s}\sqrt P {\bf{s'}} + {{{\bf{n'}}}_s}} \right) \ge \sigma _s^2\kappa } \right\} \notag\\
 &= \text{Pr}\left\{ {{{\left| {{h_s}\sqrt P {{s'}_1} + {{n'}_{s,1}}} \right|}^2} \ge \sigma _s^2\kappa } \right\} \notag \\
 &= {Q_1}\left( {\sqrt {\frac{2}{{\sigma _s^2}}P{{\left\| {{h_s}{\bf{s}}} \right\|}^2}} ,\sqrt {2\kappa } } \right) \notag \\
 &\mathop \simeq \limits^{T \gg 1} {Q_1}\left( {\sqrt {\frac{2}{{\sigma _s^2}}PT{{\left| {{h_s}} \right|}^2}} ,\sqrt {2\kappa } } \right), \label{eq19}
\end{align}where we have ${\bf{s'}} = {{\bf{U}}^H}{\bf{s}}$.

Observing from \eqref{eq18} and \eqref{eq19} that the PFA is independent of the transmit power due to the signal-free GLRT function under the $\mathcal{H}_0$ hypothesis, while the PD increases with the transmit power and the signal length due to the monotonicity of the Marcum $\text{Q}$-function with respect to (\emph{w.r.t.}) its non-centrality parameter.
\paragraph{${\mathbf{s}}_{c}$ estimated at the SR}
As stated earlier, in a general ISAC system, the communication signal, ${\mathbf{s}}_{c}$, cannot be always reconstructed accurately from the received waveform due to the noise at the SR and the channel mismatches. In the following, we will consider a scenario where the communication waveform ${{\mathbf{s}}_c}$ is substituted by its estimated version from the received ISAC signals, which doubtlessly serves as a lower bound for the communication-assisted sensing scenario. Specifically, the signals received at the SR under the null hypothesis and the alternative hypothesis are expressed by
\begin{align}
\left\{ {\begin{array}{*{20}{l}}
{{{{\bf{\tilde r}}}_s} = {{{\bf{\tilde n}}}_s},}&{{\mathcal{H}_0}},\\
{{{{\bf{\tilde r}}}_s} = \sqrt {{\rho _s}P} {h_s}{{\bf{s}}_s} + \sqrt {{\rho _c}P} {h_s}{{\bf{s}}_c} + {{{\bf{\tilde n}}}_s},}&{{\mathcal{H}_1}},
\end{array}} \right.
 \end{align}respectively.

Given a tentative value of $h_s$, the maximum likelihood estimate of ${{\bf{s}}_c}$ can be obtained by
\begin{align}
\hat{\mathbf{s}}_{c} &= \arg \mathop {\max }\limits_{{{\bf{s}}_c}} \ln f\left( {{{{\bf{\tilde r}}}_s}\left| {{\mathcal{H}_1}} \right.} \right) \notag \\
&= \arg \mathop {\min }\limits_{{{\bf{s}}_c}} {\left\| {{{{\bf{\tilde r}}}_s} - \sqrt {{\rho _s}P} {h_s}{{\bf{s}}_s} - \sqrt {{\rho _c}P} {h_s}{{\bf{s}}_c}} \right\|^2} \notag\\
&= \frac{{h_s^*}}{{\sqrt {{\rho _c}P} {{\left| {{h_s}} \right|}^2}}}{{\bf{\tilde r}}_s} - \sqrt {\frac{{{\rho _s}}}{{{\rho _c}}}} {{\bf{s}}_s}. \label{eq21}
\end{align}

Upon substituting \eqref{eq21} into \eqref{eq12}, the logarithmic GLRT function degrades into
\begin{align}
\ln \Lambda \left( {{{{\bf{\tilde r}}}_s}} \right) = \frac{1}{{\sigma _s^2}}{\left\| {{{{\bf{\tilde r}}}_s}} \right\|^2}, \label{eq22}
\end{align}which is shown to be independent of the channel coefficient $h_s$. Hence, the logarithmic GLRT function for the communication-assisted sensing case with unknown $h_s$ is the same as \eqref{eq22}.

Note that \eqref{eq22} turns out to be the classic energy detector \cite{Book_2013_Poor_An}, which also corresponds to the scenario without compensating for the delay and Doppler shift due to the fact that we have $\left \| \tilde{\mathbf{r}}_{s} \right \|^{2}=\left \| \mathbf{r}_{s} \right \|^{2}$. Consequently, the PFA in this case can be written as
\begin{align}
{P_{\textrm{FA}}} &= \text{Pr}\left\{ {\left. {\frac{1}{{\sigma _s^2}}{\left\| {{{{\bf{\tilde r}}}_s}} \right\|^2} \ge \kappa} \right|{\mathcal{H}_0}} \right\} \notag\\
&= \text{Pr}\left\{ {{{\left\| {\frac{{\sqrt 2 }}{{{\sigma _s}}}{{{\bf{\tilde n}}}_s}} \right\|}^2} \ge 2\kappa } \right\} = {Q_T}\left( {0,\sqrt {2\kappa } } \right). \label{eq23}
\end{align}

Similarly, the PD is thus expressed as
\begin{align}
{P_{\textrm{D}}} &= \text{Pr}\left\{ {\left. {\frac{1}{{\sigma _s^2}}{\left\| {{{{\bf{\tilde r}}}_s}} \right\|^2} \ge \kappa } \right|{\mathcal{H}_1}} \right\} \notag \\
&= \text{Pr}\left\{ {{{\left\| {\frac{{\sqrt {2P} }}{{{\sigma _s}}}{h_s}{\bf{s}} + \frac{{\sqrt 2 }}{{{\sigma _s}}}{{{\bf{\tilde n}}}_s}} \right\|}^2} \ge 2\kappa } \right\} \notag \\
 &= {Q_T}\left( {\sqrt {\frac{{2P}}{{\sigma _s^2}}{{\left\| {{h_s}{\bf{s}}} \right\|}^2}} ,\sqrt {2\kappa } } \right) \notag \\
 &\mathop \simeq \limits^{T \gg 1} {Q_T}\left( {\sqrt {\frac{{2PT}}{{\sigma _s^2}}{{\left| {{h_s}} \right|}^2}} ,\sqrt {2\kappa } } \right). \label{eq24}
\end{align}

In a nutshell, by assuming that the communication waveform ${{\bf{s}}_c}$ is perfectly known at the SR, \eqref{eq15} and \eqref{eq19} characterize the upper bound of the PD in communication-assisted sensing scenarios, where the channel coefficient ${h_s}$ is known and unknown by the SR, respectively. By contrast, the PD values under both these two cases are lower bounded by \eqref{eq24}, in which the communication waveform ${{\bf{s}}_c}$ is estimated at the SR.

\subsubsection{Communication-Interfered Sensing}
Next, we consider the communication-interfered sensing scenario in which the communication signal ${{\bf{s}}_c}$ is fully unknown and thus is regarded as interference at the SR\footnote{Note that we consider the device-based sensing in this paper \cite{CST_2022_Liu_A}, thus the sensing waveform ${{\bf{s}}_s}$ is always known at the SR and thus can be exploited to perform the target detection, which is in contrast to the passive radar that fails to work when the communication signal from the reference channel is unavailable.}. In this paper, we consider the worst case by assuming that the communication signal is subject to Gaussian distribution, i.e., ${{\bf{s}}_c} \sim {\mathcal{CN}}\left( {{{\bf{0}}_{T \times 1}},{\rho _c}{{\bf{I}}_T}} \right)$. As a result, the PDF of ${{{{\bf{\tilde r}}}_s}}$ under hypotheses ${{\mathcal{H}_1}}$ turns to
\begin{align}
f\left( {{{{\bf{\tilde r}}}_s}\left| {{\mathcal{H}_1}} \right.} \right) & = \frac{1}{{{\pi ^T}{{\left( {{\rho _c}P{{\left| {{h_s}} \right|}^2} + \sigma _s^2} \right)}^T}}}{{\mathop{\rm e}\nolimits} ^{ - \frac{{{{\left\| {{{{\bf{\tilde r}}}_s} - \sqrt {{\rho _s}P} {h_s}{{\bf{s}}_s}} \right\|}^2}}}{{{\rho _c}P{{\left| {{h_s}} \right|}^2} + \sigma _s^2}}}}.
\end{align}

Next, we will examine two potential scenarios depending on whether $h_s$ is known at the SR.
\paragraph{$h_s$ known at the SR}
Considering a known channel coefficient $h_s$, the logarithmic GLRT function in \eqref{eq12} is replaced by
\begin{align}
\ln\Lambda \left( {{{{\bf{\tilde r}}}_s}} \right) =& \frac{{{{\left\| {{{{\bf{\tilde r}}}_s}} \right\|}^2}}}{{\sigma _s^2}} - \frac{{{{\left\| {{{{\bf{\tilde r}}}_s} - \sqrt {{\rho _s}P} {h_s}{{\bf{s}}_s}} \right\|}^2}}}{{{\rho _c}P{{\left| {{h_s}} \right|}^2} + \sigma _s^2}} \notag \\
&+ T\ln\frac{{\sigma _s^2}}{{{\rho _c}P{{\left| {{h_s}} \right|}^2} + \sigma _s^2}} \notag\\
 =& \frac{\zeta}{{\sigma _s^2\left( {1 + \zeta } \right)}}{\left\| {{{{\bf{\tilde r}}}_s} + \frac{{\sqrt {{\rho _s}P} }}{\zeta}{h_s}{{\bf{s}}_s} } \right\|^2} - \frac{{{\rho _s}}}{{{\rho _c}}}{\left\| {{{\bf{s}}_s}} \right\|^2} \notag \\
 &- T\ln\left( {1 + \zeta } \right), \label{eq26}
\end{align}
where we have $\zeta = \frac{{{\rho _c}P{{\left| {{h_s}} \right|}^2}}}{{\sigma _s^2}}$.

Note that ${\left\| {{{{\bf{\tilde r}}}_s} + \frac{{\sqrt {{\rho _s}P} }}{\zeta}{h_s}{{\bf{s}}_s}} \right\|^2}$ in \eqref{eq26} is a non-central chi-squared distributed variable with $2T$ degrees-of-freedom such that $\frac{2}{\sigma _{s}^{2}}{\left\| {{{{\bf{\tilde r}}}_s} + \frac{{\sqrt {{\rho _s}P} }}{\zeta}{h_s}{{\bf{s}}_s}} \right\|^2}$
\begin{align}\label{eq26_0}
 \left\{ {\begin{array}{*{20}{l}}
{ \sim \chi _{2T}^{2}\left ( \frac{2\rho _{s}}{\rho _{c}\zeta }\left \| \mathbf{s}_{s} \right \|^{2} \right ),}&{{\mathcal{H}_0}},\\
{ \sim \chi _{2T}^{2}\left (\frac{2}{\sigma _{s}^{2}} \left \| \sqrt{\rho _{c}P}h_{s}\mathbf{s}_{c}+\frac{\zeta +1}{\zeta}\sqrt{\rho _{s}P}h_{s}\mathbf{s}_{s} \right \|^{2} \right ),}&{{\mathcal{H}_1}}.
\end{array}} \right.
\end{align}
under the hypotheses ${\mathcal{H}_0}$ and ${\mathcal{H}_1}$, respectively.

Based on \eqref{eq26_0}, the PFA is thus expressed as \eqref{eq27} at the top of the next page.
\begin{figure*}
\begin{align}
{P_{\textrm{FA}}} &= \text{Pr}\left\{ {\left. {\left[ \frac{\zeta }{{\sigma _s^2\left( {1 + \zeta } \right)}}{{\left\| {{{{\bf{\tilde r}}}_s} + \frac{{\sqrt {{\rho _s}P} }}{\zeta}{h_s}{{\bf{s}}_s}} \right\|}^2} - \frac{{{\rho _s}}}{{{\rho _c}}}{{\left\| {{{\bf{s}}_s}} \right\|}^2} - T\ln\left( {1 + \zeta } \right) \right] \ge \kappa } \right|{\mathcal{H}_0}} \right\} \notag\\
&= \text{Pr}\left\{ {{{\left\| {{{{\bf{\tilde n}}}_s} + \frac{{\sqrt {{\rho _s}P} }}{\zeta}{h_s}{{\bf{s}}_s}} \right\|}^2} \ge \frac{{\sigma _s^2\left( {1 + \zeta } \right)}}{\zeta }\left( {\kappa + T\ln\left( {1 + \zeta } \right) + \frac{{{\rho _s}}}{{{\rho _c}}}{{\left\| {{{\bf{s}}_s}} \right\|}^2}} \right)} \right\}\notag\\
 &= {Q_T}\left( {\sqrt {\frac{{2{\rho _s}{{\left\| {{{\bf{s}}_s}} \right\|}^2}}}{{{\rho _c}\zeta }}} ,\sqrt {\frac{{2\left( {1 + \zeta } \right)}}{\zeta }\left( {\kappa + T\ln\left( {1 + \zeta } \right) + \frac{{{\rho _s}}}{{{\rho _c}}}{{\left\| {{{\bf{s}}_s}} \right\|}^2}} \right)} } \right) \notag\\
&\mathop \simeq \limits^{T \gg 1} {Q_T}\left( {\sqrt {\frac{{2{\rho _s}T}}{{{\rho _c}\zeta }}} ,\sqrt {\frac{{2\left( {1 + \zeta } \right)}}{\zeta }\left( {\kappa + T\ln\left( {1 + \zeta } \right) + \frac{{{\rho _s}T}}{{{\rho _c}}}} \right)} } \right). \label{eq27}
\end{align}
\hrulefill
\end{figure*}
Similarly, the PD is given by \eqref{eq28} at the top of the next page.
\begin{figure*}
\begin{align}
{P_{\textrm{D}}} &= \text{Pr}\left\{ {\left. {\left[ {\frac{\zeta }{{\sigma _s^2\left( {1 + \zeta } \right)}}{{\left\| {{{{\bf{\tilde r}}}_s} + \frac{{\sqrt {{\rho _s}P} }}{\zeta}{h_s}{{\bf{s}}_s}} \right\|}^2} - \frac{{{\rho _s}}}{{{\rho _c}}}{{\left\| {{{\bf{s}}_s}} \right\|}^2} - T\ln\left( {1 + \zeta } \right)} \right] \ge \kappa } \right|{\mathcal{H}_1}} \right\} \notag\\
 &= \text{Pr}\left\{ {{{\left\| {{{{\bf{\tilde n}}}_s} + \sqrt {{\rho _c}P} {h_s}{{\bf{s}}_c} + \frac{{\left( {1 + \zeta } \right)\sqrt {{\rho _s}P} }}{\zeta }{h_s}{{\bf{s}}_s}} \right\|}^2} \ge \frac{{\sigma _s^2\left( {1 + \zeta } \right)}}{\zeta }\left( {\kappa + T\ln\left( {1 + \zeta } \right) + \frac{{{\rho _s}}}{{{\rho _c}}}{{\left\| {{{\bf{s}}_s}} \right\|}^2}} \right)} \right\} \notag\\
 &= {Q_T}\left( {\left( {\sqrt {\frac{2}{{\sigma _s^2}}{{\left\| {\sqrt {{\rho _c}P} {h_s}{{\bf{s}}_c} + \frac{{\left( {1 + \zeta } \right)\sqrt {{\rho _s}P} }}{\zeta }{h_s}{{\bf{s}}_s}} \right\|}^2}} ,} \right.\sqrt {\frac{{2\left( {1 + \zeta } \right)}}{\zeta }\left( {\kappa + T\ln\left( {1 + \zeta } \right) + \frac{{{\rho _s}}}{{{\rho _c}}}{{\left\| {{{\bf{s}}_s}} \right\|}^2}} \right)} } \right) \notag\\
&\mathop \simeq \limits^{T \gg 1} {Q_T}\left( {\left( {\sqrt {\frac{{2T}}{{{\rho _c}\zeta }}\left( {{\rho _c}{\zeta ^2} + {\rho _s}{{\left( {1 + \zeta } \right)}^2}} \right)} ,} \right.\sqrt {\frac{{2\left( {1 + \zeta } \right)}}{\zeta }\left( {\kappa + T\ln\left( {1 + \zeta } \right) + \frac{{{\rho _s}T}}{{{\rho _c}}}} \right)} } \right). \label{eq28}
\end{align}
\hrulefill
\end{figure*}
Note that the PD increases while the PFA decreases with the growing portion of the sensing power, which will be verified in Section \ref{sec5}.
\paragraph{$h_s$ unknown at the SR}
In this case, the maximum likelihood estimate of $h_s$ can be determined by
\begin{align}
{\hat h_s} =& \arg \mathop {\min }\limits_{{h_s}} \left( {T\ln\left( {{\rho _c}P{{\left| {{h_s}} \right|}^2} + \sigma _s^2} \right) } \right. \notag \\
&\qquad \qquad + \left.{\frac{{{{\left\| {{{{\bf{\tilde r}}}_s} - \sqrt {{\rho _s}P} {h_s}{{\bf{s}}_s}} \right\|}^2}}}{{{\rho _c}P{{\left| {{h_s}} \right|}^2} + \sigma _s^2}}} \right).\label{eq29}
\end{align}

Note that for a given modulus value of ${{h_s}}$, the optimal argument that minimizes the right-hand side (RHS) of \eqref{eq29} is obtained by $\angle {\hat h_s} = \angle \left( {{\bf{s}}_s^H{{{\bf{\tilde r}}}_s}} \right)$. Denote ${h_s} = \xi {\bf{s}}_s^H{{\bf{\tilde r}}_s},\ \xi \in {{\mathbb{R}}^ + }$, thus the problem in \eqref{eq29} is transformed into \eqref{eq30}, as shown at the top of this page.
\begin{figure*}
\begin{align}
 \xi = \arg \mathop {\min }\limits_{\xi \ge 0} \left( {T\ln\left( {{\rho _c}P{\xi}^2 {{\left| {{\bf{s}}_s^H{{{\bf{\tilde r}}}_s}} \right|}^2} + \sigma _s^2} \right) + \frac{{{{\left\| {{{{\bf{\tilde r}}}_s} - \sqrt {{\rho _s}P} \xi {{\bf{s}}_s}{\bf{s}}_s^H{{{\bf{\tilde r}}}_s}} \right\|}^2}}}{{{\rho _c}P{\xi}^2 {{\left| {{\bf{s}}_s^H{{{\bf{\tilde r}}}_s}} \right|}^2} + \sigma _s^2}}} \right). \label{eq30}
\end{align}
\hrulefill
\end{figure*}

It is shown that the optimal solution of \eqref{eq30} can be achieved by taking the derivative of the RHS of \eqref{eq30} \emph{w.r.t.} $\xi$ and setting it equal to $0$, i.e.,
\begin{align}
&T\frac{{2{\rho _c}P{{\left| {{\bf{s}}_s^H{{{\bf{\tilde r}}}_s}} \right|}^2}\xi}}{{{\rho _c}P{\xi}^2 {{\left| {{\bf{s}}_s^H{{{\bf{\tilde r}}}_s}} \right|}^2} + \sigma _s^2}} \notag\\
-&\frac{2{\rho _c}P{\left| {{\bf{s}}_s^H{{{\bf{\tilde r}}}_s}} \right|^2}\xi{{{\left\| {{{{\bf{\tilde r}}}_s} - \sqrt {{\rho _s}P} \xi {{\bf{s}}_s}{\bf{s}}_s^H{{{\bf{\tilde r}}}_s}} \right\|}^2}}}{{{{\left( {{\rho _c}P{\xi}^2 {{\left| {{\bf{s}}_s^H{{{\bf{\tilde r}}}_s}} \right|}^2} + \sigma _s^2} \right)}^2}}} \notag \\
+& \frac{{2\xi {{\left\| {\sqrt {{\rho _s}P} {{\bf{s}}_s}{\bf{s}}_s^H{{{\bf{\tilde r}}}_s}} \right\|}^2} - 2\Re\left\{ {{\bf{\tilde r}}_s^H\sqrt {{\rho _s}P} {{\bf{s}}_s}{\bf{s}}_s^H{{{\bf{\tilde r}}}_s}} \right\}}}{{{\rho _c}P{\xi}^2 {{\left| {{\bf{s}}_s^H{{{\bf{\tilde r}}}_s}} \right|}^2} + \sigma _s^2}} = 0,\label{eq32}
\end{align}
which can be numerically solved using the bisection method.

Let $\hat \xi$ denote the estimated $\xi$ by solving \eqref{eq32}. Upon substituting the estimated ${\hat h_s} = \hat \xi {\bf{s}}_s^H{{\bf{\tilde r}}_s}$ into \eqref{eq26}, we arrive at
\begin{align}\label{eq34_3}
\ln\Lambda \left( {{{{\bf{\tilde r}}}_s}} \right) =& \frac{{{{\left\| {{{{\bf{\tilde r}}}_s}} \right\|}^2}}}{{\sigma _s^2}} - \frac{{{{\left\| {{{{\bf{\tilde r}}}_s} - \sqrt {{\rho _s}P} \hat \xi {{\bf{s}}_s}{\bf{s}}_s^H{{{\bf{\tilde r}}}_s}} \right\|}^2}}}{{{\rho _c}P{\hat \xi ^2}\left | \mathbf{s}_{s}^{H}\tilde{\mathbf{r}}_s \right |^{2} + \sigma _s^2}} \notag \\
&+ T\ln\frac{{\sigma _s^2}}{{{\rho _c}P{\hat \xi ^2}\left | \mathbf{s}_{s}^{H}\tilde{\mathbf{r}}_s \right |^{2} + \sigma _s^2}}.
\end{align}

Since the closed-form PDF of $\ln\Lambda \left( {{{{\bf{\tilde r}}}_s}} \right)$ in \eqref{eq34_3} is unknown due to the logarithm and division operations, in this paper, we develop a reference bound of PFA and PD by replacing $\left | \mathbf{s}_{s}^{H}\tilde{\mathbf{r}}_s \right |^{2}$ in \eqref{eq34_3} with its statistical value $\mathbb{E}\left ( \left | \mathbf{s}_{s}^{H}\tilde{\mathbf{r}}_s \right |^{2} \right )$ under each hypothesis \cite{Book_2013_Poor_An}. Note that the approximation becomes more accurate as more symbols are collected for sensing, i.e., $T\gg 1$. By defining $\vartheta \triangleq \mathbb{E}\left ( \left | \mathbf{s}_{s}^{H}\tilde{\mathbf{r}}_s \right |^{2} \right )$, we have
\begin{align}
\left\{ {\begin{array}{*{20}{l}}
{\vartheta = 0,}&{{\mathcal{H}_0}},\\
{\vartheta = \rho _{s}P\left \| \mathbf{s}_{s} \right \|^{4}\mathbb{E}\left ( \left | h_{s} \right |^{2} \right ),}&{{\mathcal{H}_1}},
\end{array}} \right.
\end{align}where the second-order statistic of channel coefficient, i.e., $\mathbb{E}\left ( \left | h_{s} \right |^{2} \right )$, is assumed to be \emph{a priori} known at the SR.

As a result, \eqref{eq34_3} can be rewritten as \eqref{eq34_4} at the top of the next page,
\begin{figure*}
\begin{align}\label{eq34_4}
\ln\Lambda \left( {{{{\bf{\tilde r}}}_s}} \right) = \frac{\tilde{\mathbf{r}}_{s}^{H}\tilde{\mathbf{r}}_{s}}{\sigma _{s}^{2}}-\frac{\tilde{\mathbf{r}}_{s}^{H}\left ( \mathbf{I}-\sqrt{\rho _{s}P}\hat{\xi }\mathbf{s}_{s}\mathbf{s}_{s}^{H} \right )^{H}\left ( \mathbf{I}-\sqrt{\rho _{s}P}\hat{\xi }\mathbf{s}_{s}\mathbf{s}_{s}^{H} \right )\tilde{\mathbf{r}}_{s}}{\rho _{c}P\hat{\xi} ^{2}\vartheta +\sigma _{s}^{2}} +T\ln\frac{\sigma _{s}^{2}}{\rho _{c}P\hat{\xi} ^{2}\vartheta +\sigma _{s}^{2}} = {\bf{\tilde r}}_s^H{\bf{K}}{{{\bf{\tilde r}}}_s} + T\ln\frac{{\sigma _s^2}}{{{\rho _c}P{\hat \xi ^2}\vartheta + \sigma _s^2}},
\end{align}
\hrulefill
\end{figure*}
where we have 
\begin{align}
{\bf{K}} =& \frac{1}{{\sigma _s^2}}{\bf{I}} - \frac{1}{{{\rho _c}P{\hat \xi ^2}\vartheta+ \sigma _s^2}}\notag \\
&\times {\left( {{\bf{I}} - \sqrt {{\rho _s}P} \hat \xi {{\bf{s}}_s}{\bf{s}}_s^H} \right)^H}\left( {{\bf{I}} - \sqrt {{\rho _s}P} \hat \xi {{\bf{s}}_s}{\bf{s}}_s^H} \right).
\end{align}

Let ${\bf{K}} = {\bf{V\Sigma }}{{\bf{V}}^H}$ denote the eigenvalue decomposition of ${\bf{K}}$. It is obvious that one eigenvalue of ${\bf{K}}$ is ${\frac{1}{{\sigma _s^2}} - \frac{1}{{{\rho _c}P{\hat \xi ^2}\vartheta + \sigma _s^2}}{{\left( {1 - \sqrt {{\rho _s}P} \hat \xi {\bf{s}}_s^H{{\bf{s}}_s}} \right)}^2}}$, while the remaining $\left( {T - 1} \right)$ eigenvalues are $\left( {\frac{1}{{\sigma _s^2}} - \frac{1}{{{\rho _c}P{\hat \xi ^2}\vartheta + \sigma _s^2}}} \right)$.

By denoting ${{{\tilde{{\mathbf{n}}}}''}_{s}}=\mathbf{V}^{H}\tilde{\mathbf{n}}_{s}$, we have ${{\mathbf{n}}''}_{s} \sim {\mathcal{CN}}\left( {{{\bf{0}}_{T \times 1}},{\sigma _s^2}{{\bf{I}}_T}} \right)$. As a result, the PFA can be expressed as
\begin{align}
{P_{\textrm{FA}}} &= \text{Pr}\left\{ {\left. {{\bf{\tilde r}}_s^H{\bf{K}}{{{\bf{\tilde r}}}_s} + T\ln\frac{{\sigma _s^2}}{{{\rho _c}P{\hat \xi ^2}\vartheta + \sigma _s^2}} \ge \kappa } \right|{\mathcal{H}_0}} \right\} \notag \\
&= \text{Pr}\left \{ {{\mathbf{n}}''}_{s}^{H}\mathbf{\Sigma}{{\mathbf{n}}''}_s \geq \kappa -T\log\frac{\sigma _{s}^{2}}{\rho P\hat{\xi}^{2}\vartheta + \sigma _{s}^{2}} \right \} \notag \\
&=\text{Pr}\left\{ {\sum\limits_{l = 2}^T {{{\left| {\frac{{\sqrt 2 }}{{{\sigma _s}}}{{n}_{s,l}''}} \right|}^2}} \ge \alpha - \beta{{\left| {\frac{{\sqrt 2 }}{{{\sigma _s}}}{{n}_{s,1}''}} \right|}^2}} \right\}, \label{eq34}
\end{align}where we have 
\begin{align}
\alpha &= 2\left( {\kappa - T\ln\frac{{\sigma _s^2}}{{{\rho _c}P{\hat \xi ^2}\vartheta + \sigma _s^2}}} \right)\left( {1 + \frac{{\sigma _s^2}}{{{\rho _c}P{\hat \xi ^2}\vartheta}}} \right),\\
\beta &= 1 + \frac{{\sigma _s^2}}{{{\rho _c}P{\hat \xi ^2}\vartheta}} - \frac{{\sigma _s^2}}{{{\rho _c}P{\hat \xi ^2}\vartheta}}{\left( {1 - \sqrt {{\rho _s}P} \hat \xi {\bf{s}}_s^H{{\bf{s}}_s}} \right)^2}. \label{eq34_1}
\end{align}

Furthermore, defining ${X_0} \buildrel \Delta \over = \sum\limits_{l = 2}^T {{{\left| {\frac{{\sqrt 2 }}{{{\sigma _s}}}{{n''}_{s,l}}} \right|}^2}} $ and ${Y_0} \buildrel \Delta \over = {\left| {\frac{{\sqrt 2 }}{{{\sigma _s}}}{{n''}_{s,1}}} \right|^2}$, we have ${X_0}\sim \chi _{2\left( {T - 1} \right)}^2\left( 0 \right)$ and ${Y_0} \sim \chi _2^2\left( 0 \right)$. As a result, \eqref{eq34} can be rewritten as
\begin{align}
{P_{\textrm{FA}}} =& \text{Pr}\left\{ {{X_0} \ge \alpha - \beta {Y_0}} \right\} \notag \\
=& 1 - \frac{1}{2}\int_0^{\frac{\alpha}{\beta }} {\frac{{\gamma \left( {T - 1,\frac{{\alpha - \beta {y_0}}}{2}} \right)}}{{\Gamma \left( {T - 1} \right)}}{e^{ - \frac{{y_0}}{2}}}d{y_0}} \notag\\
 \overset{\left( a \right)}{=}& 1 - \frac{1}{2}\int_0^{\frac{\alpha}{\beta }} {\left( {1 - {e^{ - \frac{{(\alpha - \beta {y_0})}}{2}}}(\sum\limits_{l = 0}^{T - 2} {\frac{{{{\left( {\alpha - \beta {y_0}} \right)}^l}}}{{{2^l}l!}}} )} \right)} \notag \\
 & \times {e^{ - \frac{{y_0}}{2}}}d{y_0} \notag\\
 =& {e^{ - \frac{\alpha}{{2\beta }}}} + \frac{1}{2}{e^{ - \frac{\alpha}{2}}}\sum\limits_{l = 0}^{T - 2} {\frac{{{\beta ^l}\int_0^{\frac{\alpha}{\beta }} {{{\left( {\frac{\alpha}{\beta } - {y_0}} \right)}^l}{e^{ - \left( {\frac{{1 - \beta }}{2}} \right){y_0}}}d{y_0}} }}{{{2^l}l!}}} \notag \\
 \overset{\left( b \right)}{=}&  \left( {1 + \frac{1}{\beta }\sum\limits_{l = 1}^{T - 1} {\frac{{{\beta ^l}}}{{{{\left( {\beta - 1} \right)}^l}\left( {l - 1} \right)!}}} \gamma \left( {l,\frac{\alpha}{{2\beta }}\left( {\beta - 1} \right)} \right)} \right) \notag \\
 &\times {e^{ - \frac{\alpha}{{2\beta }}}}, \label{eq35}
\end{align}
where $\left( a \right)$ and $\left( b \right)$ hold by applying \emph{Eq. (8.352)} and \emph{Eq. (3.382)}, respectively, of \cite{Book_2014_Gradshteyn_Table}.

Similarly, by defining ${{\mathbf{s}}''}=\mathbf{V}^{H}\mathbf{s}$, the PD is given by \eqref{eq36} at the top of the next page.
\begin{figure*}
\begin{align}
{P_{\textrm{D}}} &= \text{Pr}\left\{ {\left. {{\bf{\tilde r}}_s^H{\bf{K}}{{{\bf{\tilde r}}}_s} + T\ln\frac{{\sigma _s^2}}{{{\rho _c}P{\hat \xi ^2}\vartheta + \sigma _s^2}} \ge \kappa } \right|{\mathcal{H}_1}} \right\} \notag\\
&= \text{Pr}\left\{ {\sum\limits_{l = 2}^T {{{\left| {\frac{{\sqrt {2P} {h_s}{{s''}_l}}}{{{\sigma _s}}} + \frac{{\sqrt 2 }}{{{\sigma _s}}}{{n''}_{s,l}}} \right|}^2}} \ge \alpha - \beta {{\left| {\frac{{\sqrt {2P} {h_s}{{s''}_1}}}{{{\sigma _s}}} + \frac{{\sqrt 2 }}{{{\sigma _s}}}{{n''}_{s,1}}} \right|}^2}} \right\}. \label{eq36}
\end{align}
\hrulefill
\end{figure*}
Upon defining
\begin{align}
    {X_1} \buildrel \Delta \over =& \sum\limits_{l = 2}^T {{{\left| {\frac{{\sqrt {2P} {h_s}{{s''}_l}}}{{{\sigma _s}}} + \frac{{\sqrt 2 }}{{{\sigma _s}}}{{n''}_{s,l}}} \right|}^2}},\\
    {Y_1} \buildrel \Delta \over =& {\left| {\frac{{\sqrt {2P} {h_s}{{s''}_1}}}{{{\sigma _s}}} + \frac{{\sqrt 2 }}{{{\sigma _s}}}{{n''}_{s,1}}} \right|^2},
\end{align}
we have ${X_1}\sim \chi _{2\left( {T - 1} \right)}^2\left( {\frac{{2P\left( {T - 1} \right){{\left\| {{h_s}}{\bf{s}} \right\|}^2}}}{{T\sigma _s^2}}} \right)$ and ${Y_1}\sim \chi _2^2\left( {\frac{{2P{{\left\| {{h_s}}{\bf{s}} \right\|}^2}}}{{T\sigma _s^2}}} \right)$. Thus, \eqref{eq36} can be rewritten as \eqref{eq37} at the top of the next page.
\begin{figure*}
\begin{align}
{P_{\textrm{D}}} &= \text{Pr}\left\{ {{X_1} \ge \alpha - \beta {Y_1}} \right\} \notag \\
&= 1 - \frac{1}{2}{e^{ - \frac{{P{{\left\| {{h_s}}{\bf{s}} \right\|}^2}}}{{T\sigma _s^2}}}}\int_0^{\frac{\alpha}{\beta }} {{e^{ - \frac{{{y_1}}}{2}}}{I_0}\left( {\sqrt {\frac{{2P{{\left\| {{h_s}}{\bf{s}} \right\|}^2}}}{{T\sigma _s^2}}{y_1}} } \right) \left( {1 - {Q_{T - 1}}\left( {\sqrt {\frac{{2P\left( {T - 1} \right){{\left\| {{h_s}}{\bf{s}} \right\|}^2}}}{{T\sigma _s^2}}} ,\sqrt {\left( {\alpha - \beta {y_1}} \right)} } \right)} \right)d{y_1}}. \label{eq37}
\end{align}
\hrulefill
\end{figure*}

Since it is non-trivial to obtain a closed-form expression for \eqref{eq37}, we next provide an approximated expression by applying Chebyshev-Gaussian quadrature \cite{Book_1988_Abramowitz_Handbook}. Specifically, by dividing the integral domain into $N$ parts and defining $y_{1}=\frac{\alpha }{2\beta }\left ( 1+\cos\left ( \frac{2n-1}{2N}\pi \right ) \right ),\ n=1,2,\cdots ,N$, the PD can be numerically approximated by \eqref{eq38} at the top of the next page,
\begin{figure*}
\begin{align}
{P_{\textrm{D}}} \approx& 1 - \frac{{\pi \alpha}}{{4\beta N}}{e^{ - \frac{{P{{\left\| {{h_s}}{\bf{s}} \right\|}^2}}}{{T\sigma _s^2}} - \frac{\alpha}{{4\beta }}}}\left[ {\sum\limits_{n = 1}^N {\sin \left( {\frac{{2n - 1}}{{2N}}\pi } \right){e^{ - \frac{\alpha}{{4\beta }}\cos \left( {\frac{{2n - 1}}{{2N}}\pi } \right)}}} } {I_0}\left( {\sqrt {\frac{{\alpha P{{\left\| {{h_s}}{\bf{s}} \right\|}^2}}}{{\beta T\sigma _s^2}}\left( {1 + \cos \left( {\frac{{2n - 1}}{{2N}}\pi } \right)} \right)} } \right) \right. \notag\\
& \left. { \times \left( {1 - {Q_{T - 1}}\left( {\sqrt {\frac{{2P\left( {T - 1} \right){{\left\| {{h_s}}{\bf{s}} \right\|}^2}}}{{T\sigma _s^2}}} ,\sqrt {\frac{\alpha}{2}\left( {1 - \cos \left( {\frac{{2n - 1}}{{2N}}\pi } \right)} \right)} } \right)} \right)} \right], \label{eq38}
\end{align}
\hrulefill
\end{figure*}
where $N$ is an adjustable parameter to strike a flexible tradeoff between the computational complexity and the fitting accuracy. Specifically, $N$ denotes the number of pieces dividing the integral interval. Upon increasing the value of $N$ one could narrow the gap between the numerically approximated results of \eqref{eq38} and the true value of \eqref{eq36}.

When adequate symbols are collected for sensing, i.e., $T \gg 1$, \eqref{eq35} and \eqref{eq38} can be further simplified by replacing $\left \| \mathbf{s} \right \|^{2}$ with $T$. Note that due to the statistical approximation, \eqref{eq35} and \eqref{eq38} serve as a lower bound for the PFA and the PD, respectively, which will be verified later.

\section{Tradeoff Analysis via Power Allocation}\label{sec4}
In this section, we elaborate on the performance tradeoff between the sensing and communication functionalities of an ISAC system by solving the power allocation problem. Specifically, the achievable rate of the CU and the PD for target detection are highly dependent on the power allocated to the corresponding signal components. Therefore, we consider a general optimization objective of minimizing the transmit power at the ISAC-BS while ensuring that both the information rate of the CU and the PD for monitoring the target of interest at the SR are above some preset threshold values. As such, the optimization problem can be expressed as\footnote{Note that although we consider the power minimization problem in this paper, the formulated problem can be readily transformed into its dual problem of maximizing the achievable rate or PD by setting a fixed value of transmit power, which corresponds to the communication-centric and sensing-centric ISAC scenarios, respectively \cite{CST_2022_Liu_A}.}
\begin{subequations}\label{eq39}
\begin{alignat}{3}
{\left( {P1} \right)}: \ &{\mathop {\min }\limits_{\rho_c, \rho_s} } \ &&{P} \label{eq39A}\\
{}\ &{\text{s.t.}} \ &&{R \ge {R_{\min}}}, \label{eq39B}\\
{}\ &{} \ &&{{P_{\textrm{D}}} \ge P_{\textrm{D}, \min}}, \label{eq39C}\\
{}\ &{} \ &&{{P_{\textrm{FA}}} \le P_{\textrm{FA}, \delta}}, \label{eq39D}\\
{}\ &{} \ &&{\rho_c + \rho_s = 1}, \label{eq39E}\\
{}\ &{} \ &&{\rho_c\geq0,\ \rho_s\geq0,} \label{eq39F}
\end{alignat}
\end{subequations}
where ${R_{\min }}$ and ${P_{\textrm{D}, \min }}$ denote the constant minimum information rate required by the CU and the PD required for sensing, respectively; ${P_{\textrm{FA},\delta }}$ denotes the maximum tolerable value of the PFA. Next, we will solve $\left( {P1} \right)$ by considering eight different cases, which are based on the combination of the two types of communication scenarios (i.e., sensing-free and sensing-interfered) and the two types of sensing scenarios (i.e., communication-assisted and communication-interfered) discussed in Section \ref{sec3}. In each scenario, we will consider two cases with known and unknown $h_s$, respectively.

\emph{\textbf{Case I:} Sensing-Free Communication and Communication-Assisted Sensing ($\mathbf{s}_{s}$ known at CU, both $\mathbf{s}_{c}$ and $h_s$ known at SR)}

Upon combining \eqref{eq4} and \eqref{eq15}, the optimal power allocation solution for $\left( {P1} \right)$ is readily obtained as ${\rho _c} = 1$, indicating that the total transmit power should be allocated for communication. This is due to the fact that the communication signal in this case can always be fully exploited for enhancing the target detection performance. By doing so, the minimum amount of transmit power required at the ISAC-BS for satisfying \eqref{eq39B} is obtained by $\frac{\sigma _{c}^{2}}{\left | h_{c} \right |^{2}}\left ( 2^{R_{\min}} -1\right )$. Furthermore, in order to meet the sensing requirements of \eqref{eq39C} and \eqref{eq39D} with the minimum transmit power, we have $Q\left( {\frac{{\sigma _s^2\kappa + PT{{\left| {{h_s}} \right|}^2}}}{{\sqrt {2PT} {\sigma _s}\left| {{h_s}} \right|}}} \right)=P_{\text{FA},\delta }$ and $Q\left( {\frac{{\sigma _s^2\kappa - PT{{\left| {{h_s}} \right|}^2}}}{{\sqrt {2PT} {\sigma _s}\left| {{h_s}} \right|}}} \right)=P_{\text{D},\min }$. Hence, the minimum amount of transmit power required for satisfying \eqref{eq39C} and \eqref{eq39D} is attained by $\frac{{\sigma _s^2}}{{2T{{\left| {{h_s}} \right|}^2}}}{{\left[ {{Q^{ - 1}}\left( {{P_{\textrm{FA},\delta }}} \right) - {Q^{ - 1}}\left( {{P_{\textrm{D},\min }}} \right)} \right]}^2}$, where $Q^{-1}\left ( \cdot \right )$ denotes the inverse $\text{Q}$-function. As a result, the minimum amount of transmit power is given by \eqref{eq40} at the top of the next page.
\begin{figure*}
\begin{align}
{P_{\min }} = \max \left( {\frac{{\sigma _s^2}}{{2T{{\left| {{h_s}} \right|}^2}}}{{\left[ {{Q^{ - 1}}\left( {{P_{\textrm{FA},\delta }}} \right) - {Q^{ - 1}}\left( {{P_{\textrm{D},\min }}} \right)} \right]}^2},\frac{{\sigma _c^2}}{{{{\left| {{h_c}} \right|}^2}}}\left( {{2^{{R_{\min }}}} - 1} \right)} \right).\label{eq40}
\end{align}
\hrulefill
\end{figure*}

\emph{\textbf{Case II:} Sensing-Free Communication and Communication-Assisted Sensing ($\mathbf{s}_{s}$ known at CU, $\mathbf{s}_{c}$ known while $h_s$ unknown at SR)}

Similarly, the optimal power allocation solution in this case is ${\rho _c} = 1$, bearing in mind that the superimposed ISAC waveform can always be employed for carrying out the target detection. Upon combining \eqref{eq18} and \eqref{eq19}, the minimum amount of transmit power for meeting the sensing requirements of \eqref{eq39C} and \eqref{eq39D} is obtained by solving ${Q_1}\left( {\sqrt {\frac{2}{{\sigma _s^2}}PT{{\left| {{h_s}} \right|}^2}} ,\sqrt {2\kappa } } \right)={P_{\textrm{D},\min }}$ while satisfying ${e^{ - \kappa }}=P_{\text{FA},\delta }$. Let ${{P_{S,\min }}}$ denote the minimum amount of transmit power required for meeting these sensing requirements, which can be acquired by numerically solving ${Q_1}\left( {\sqrt {\frac{2}{{\sigma _s^2}}{P_{S,\min }}T{{\left| {{h_s}} \right|}^2}} ,\sqrt { - 2\ln{P_{\textrm{FA},\delta }}} } \right) = {P_{\textrm{D},\min }}$. Hence, the minimum amount of transmit power required at the ISAC-BS is given by
\begin{align}
{P_{\min }} = \max \left( {{P_{S,\min }},\frac{{\sigma _c^2}}{{{{\left| {{h_c}} \right|}^2}}}\left( {{2^{{R_{\min }}}} - 1} \right)} \right). \label{eq41}
\end{align}

In the worst case of recovering communication signals, the PFA and PD are characterized by \eqref{eq23} and \eqref{eq24}, respectively. Thus, the minimum transmit power of \emph{Cases I \& II} can be obtained upon replacing ${{P_{S,\min }}}$ in \eqref{eq41} by numerically solving ${{Q_T}\left( {\sqrt {\frac{{2P_{S,\min }T}}{{\sigma _s^2}}{{\left| {{h_s}} \right|}^2}} ,\sqrt {2\kappa } } \right) = {P_{\textrm{D},\min }}}$ via the bisection searching method, while ${\kappa }$ satisfying ${{Q_T}\left( {0,\sqrt {2\kappa } } \right) = {P_{\textrm{FA},\delta }}}$.

\emph{\textbf{Case III:} Sensing-Free Communication and Communication-Interfered Sensing ($\mathbf{s}_{s}$ known at CU, $\mathbf{s}_{c}$ unknown while $h_s$ known at SR)}

Considering the case that the communication signal acts as interference during the sensing procedure, it is evident that the optimal power allocation solution always satisfies $R = {R_{\min }}$. Thus, we have
\begin{align}
{\rho _c}{P_{\min }} = \frac{{\left( {{2^{{R_{\min }}}} - 1} \right)\sigma _c^2}}{{{{\left| {{h_c}} \right|}^2}}}. \label{eq42_0}
\end{align}

Upon substituting \eqref{eq42_0} into \eqref{eq27} and \eqref{eq28}, one could obtain the sensing power ${P_{S,\min }} = {\rho _s}{P_{\min }}$ required for satisfying \eqref{eq39C} and \eqref{eq39D}. More specifically, ${P_{S,\min }}$ is numerically calculated by solving the problem $P_{\text{D}}=P_{\text{D},\min}$, while satisfying $P_{\text{FA}}=P_{\text{FA},\delta }$, in which $P_{\text{FA}}$ and $P_{\text{D}}$ are characterized by \eqref{eq27} and \eqref{eq28}, respectively.

Furthermore, by recalling the fact that ${\rho_c + \rho_s = 1}$, the optimal power allocation solution and the minimum amount of transmit power are given by
\begin{align}
{\rho _c} &= \frac{{\left( {{2^{{R_{\min }}}} - 1} \right)\sigma _c^2}}{{{P_{S,\min }}{{\left| {{h_c}} \right|}^2} + \left( {{2^{{R_{\min }}}} - 1} \right)\sigma _c^2}}, \label{eq42}\\
{P_{\min }} &= \frac{{\left( {{2^{{R_{\min }}}} - 1} \right)\sigma _c^2}}{{{{\left| {{h_c}} \right|}^2}}} + {P_{S,\min }}, \label{eq43}
\end{align}
respectively. From \eqref{eq42} one may note that $\rho_c$ increases with $R_{\min}$ but decreases with $P_{S,\min }$. This implies that in order to meet a higher communication QoS requirement, more power should be allocated to the CU. Conversely, if there is a higher demand for sensing performance, then less power should be allocated to the CU.

\emph{\textbf{Case IV:} Sensing-Free Communication and Communication-Interfered Sensing ($\mathbf{s}_{s}$ known at CU, both $\mathbf{s}_{c}$ and $h_s$ unknown at SR)}

In this case, the sensing power required ${P_{S,\min }}$ to meet the corresponding sensing requirements, i.e., \eqref{eq39C} and \eqref{eq39D}, is obtained by solving the problem of $P_{\text{D}}=P_{\text{D},\min}$ given the target PFA value quantified by $P_{\text{FA}}=P_{\text{FA},\delta }$, where $P_{\text{FA}}$ and $P_{\text{D}}$ are characterized by \eqref{eq35} and \eqref{eq38}, respectively. Once the required sensing power has been determined, the optimal power allocation solution and the minimum transmit power are obtained by applying \eqref{eq42} and \eqref{eq43}, respectively.

\emph{\textbf{Case V:} Sensing-Interfered Communication and Communication-Assisted Sensing ($\mathbf{s}_{s}$ unknown at CU, both $\mathbf{s}_{c}$ and $h_s$ known at SR)}

With prior knowledge of the communication signal, the SR is capable of employing the whole amount of power for sensing. Specifically, the PFA in \eqref{eq14} and the PD in \eqref{eq15} are independent of $\rho_c$ and $\rho_s$, while the achievable rate characterized by \eqref{eq5} gradually increases with $\rho_c$. Therefore, the optimal power allocation solution is achieved at $\rho_c = 1$, and thus the minimum amount of transmit power required in this case is given by \eqref{eq40}, which is the same as \emph{Case I}.

\emph{\textbf{Case VI:} Sensing-Interfered Communication and Communication-Assisted Sensing ($\mathbf{s}_{s}$ unknown at CU, $\mathbf{s}_{c}$ known while $h_s$ unknown at SR)}

Similarly, due to the fact that the PFA and the PD characterized by \eqref{eq18} and \eqref{eq19} are independent on $\rho_c$ and $\rho_s$, the achievable rate in \eqref{eq5} is maximized at $\rho_c = 1$. Following the same consideration, the optimal power allocation solution in this case is the same as \emph{Case II}, while the minimum amount of transmit power is given by \eqref{eq41}.

\emph{Remark 2:} The results that the optimal power allocation solutions for \emph{Cases V and VI} are the same as those for \emph{Cases I and II} are consistent with our intuition since the communication-assisted sensing mode is considered in these four cases. As such, the communication signal can always be utilized for assisting in the target detection at the SR. Note that the optimal solutions for all these four cases are achieved at ${\rho _c} = 1$, which is equivalent to the classic passive radar without transmitting the additional sensing signal from the ISAC-BS \cite{SPL_2017_Chalise_Performance, TSP_2013_Palmer_DVB}. In a nutshell, an ISAC waveform with only the communication component proves to be the most energy-efficient solution for the communication-assisted sensing mode upon minimizing the interference at the CU.

\emph{\textbf{Case VII:} Sensing-Interfered Communication and Communication-Interfered Sensing ($\mathbf{s}_{s}$ unknown at CU, $\mathbf{s}_{c}$ unknown while $h_s$ known at SR)}

Since both sensing and communication signals in this case are regarded as interference to each other's process, it becomes clear that the optimal power allocation solution attains at $R = {R_{\min }}$. Thus, we have
\begin{align}
 {\rho _c}{P_{\min }} = \frac{{\left( {{2^{{R_{\min }}}} - 1} \right)\left( {{\rho _s}P{{\left| {{h_c}} \right|}^2} + \sigma _c^2} \right)}}{{{{\left| {{h_c}} \right|}^2}}}. \label{eq44}
\end{align}

Upon substituting \eqref{eq44} into \eqref{eq27} and \eqref{eq28} and considering that $P_{\text{D}}=P_{\text{D},\min}$ and $P_{\text{FA}}=P_{\text{FA},\delta }$, one could calculate the minimum amount of sensing power ${P_{S,\min }} = {\rho _s}{P_{\min }}$. Thus, the optimal power allocation solution and the minimum transmit power are determined by
\begin{align}
{\rho _c} &= \frac{{\left( {{2^{{R_{\min }}}} - 1} \right)\left( {{P_{S,\min }}{{\left| {{h_c}} \right|}^2} + \sigma _c^2} \right)}}{{\left( {{2^{{R_{\min }}}} - 1} \right)\left( {{P_{S,\min }}{{\left| {{h_c}} \right|}^2} + \sigma _c^2} \right) + {{\left| {{h_c}} \right|}^2}{P_{S,\min }}}}, \label{eq45}\\
{P_{\min }} &= \left( {{2^{{R_{\min }}}} - 1} \right)\frac{{\sigma _c^2}}{{{{\left| {{h_c}} \right|}^2}}} + {2^{{R_{\min }}}}{P_{S,\min }}, \label{eq46}
\end{align}
respectively.

\emph{\textbf{Case VIII:} Sensing-Interfered Communication and Communication-Interfered Sensing ($\mathbf{s}_{s}$ unknown at CU, both $\mathbf{s}_{c}$ and $h_s$ unknown at SR)}

Similarly, the optimal power allocation solution and the minimum transmit power can be obtained by applying \eqref{eq45} and \eqref{eq46}, respectively, in which ${P_{S,\min }}$ is obtained by numerically solving $P_{\textrm{D}}=P_{\textrm{D},\min}$, while satisfying $P_{\textrm{FA}}=P_{\textrm{FA},\delta }$, where $P_{\textrm{FA}}$ and $P_{\textrm{D}}$ are characterized by \eqref{eq35} and \eqref{eq38}, respectively.

In the preceding discussion, we examined the tradeoff between sensing and communication functionalities in eight typical scenarios. We note that in some scenarios, such as the communication-assisted sensing scenario, these two functionalities could achieve mutual benefits, where the communication rate could be improved without sacrificing the sensing performance. As such, both the sensing and communication performance benefit from increasing the power of the communication signal, thus achieving the optimal operating state at $\rho _{c}=1$. In the communication-interfered sensing scenario, these two functionalities behave competitively. Hence, one has to sophisticatedly perform the power allocation to minimize the mutual interference between these two components. The tradeoff analysis in this section would be quantitatively verified in Section \ref{sec5}.

\section{Simulation Results}\label{sec5}
\subsection{Simulation Setup}
\begin{figure}[!t]
\centering
\includegraphics[width=8cm]{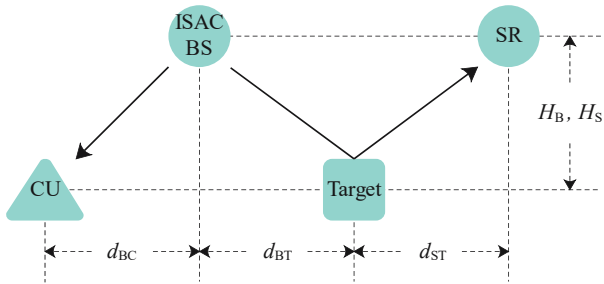}
\caption{Simulation setup of the considered ISAC system.}
\label{fig2}
\end{figure}
In this section, numerical experiments are provided to verify our theoretical analysis and evaluate the fundamental tradeoff between the PD and achievable rate in the considered ISAC system. The simulation setup is illustrated in Fig. \ref{fig2}, in which the horizontal distances between the ISAC-BS and the CU, between the ISAC-BS and the target, and between the SR and the target are all set to $d_{\textrm{BC}}=d_{\textrm{BT}}=d_{\textrm{ST}}=100$ meters (m). The heights of the ISAC-BS and the SR are both set to ${H_{\textrm{B}}} = {H_{\textrm{S}}} = 10$ m, while the CU and target are assumed to be at ground level with an altitude of $0$ m. In our simulations, we adopt the COST Hata model to characterize the path loss \cite{Book_2005_Tse_Fundamentals}, i.e.,
\begin{align}
 PL =& \left( {44.9 - 6.55{{\log }_{10}}{h_{t}}} \right){\log _{10}}d - \left( {1.1{{\log }_{10}}f - 0.7} \right){h_{r}} \notag \\
 & + 5.83{\log _{10}}{h_{t}}+ 35.46{\log _{10}}f- 89.2 \quad \text{dB},
\end{align}
where $h_{t}$ and $h_{r}$ denote the heights of transmit and receive antennas, respectively, of the corresponding link, $f$ (MHz) denotes the carrier frequency, which is set to $f = 2000$ in our ISAC system, $d$ (m) denotes the link distance, which can be easily calculated according to the geometrical layout shown in Fig. \ref{fig2}. The communication link is assumed to experience Rayleigh fading, while the sensing link is modeled by a channel coefficient determined by the propagation distance. The receiver sensitivity at the SR is set to $\sigma _s^2 = - 175$ dBm, while the noise power at the CU is set to $\sigma _c^2 = - 115$ dBm \cite{TSP_2019_Wang_Power}. All simulation results are achieved by averaging over 10,000 independent experiments.
\begin{figure*}[!t]
\centering
\subfloat[$P_{\textrm{FA}}$ or $P_{\textrm{D}}$ versus $\kappa$ with known $h_s$.]{\includegraphics[width=7.5cm]{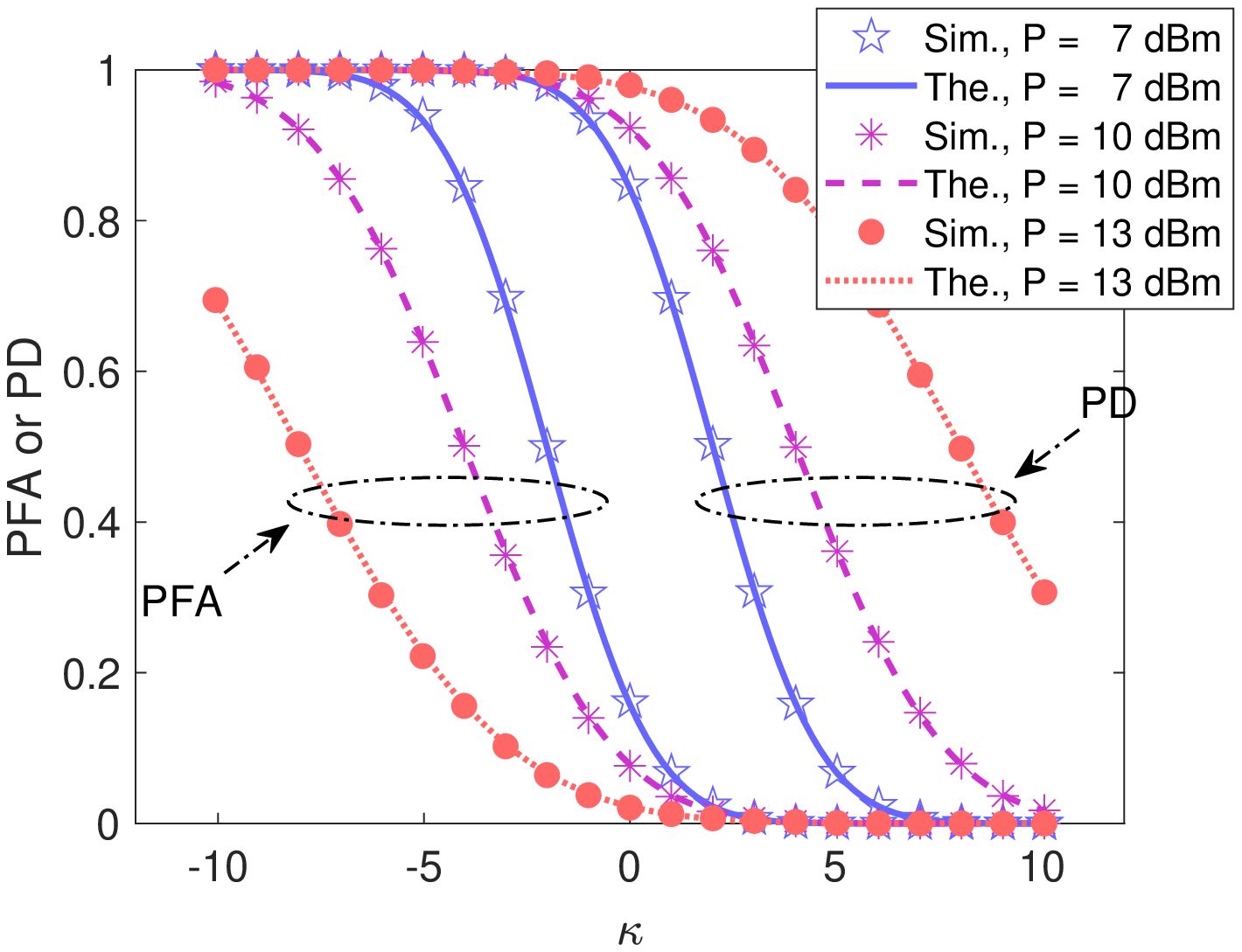}
\label{fig3a}}
\subfloat[$P_{\textrm{FA}}$ or $P_{\textrm{D}}$ versus $\kappa$ with unknown $h_s$.]{\includegraphics[width=7.5cm]{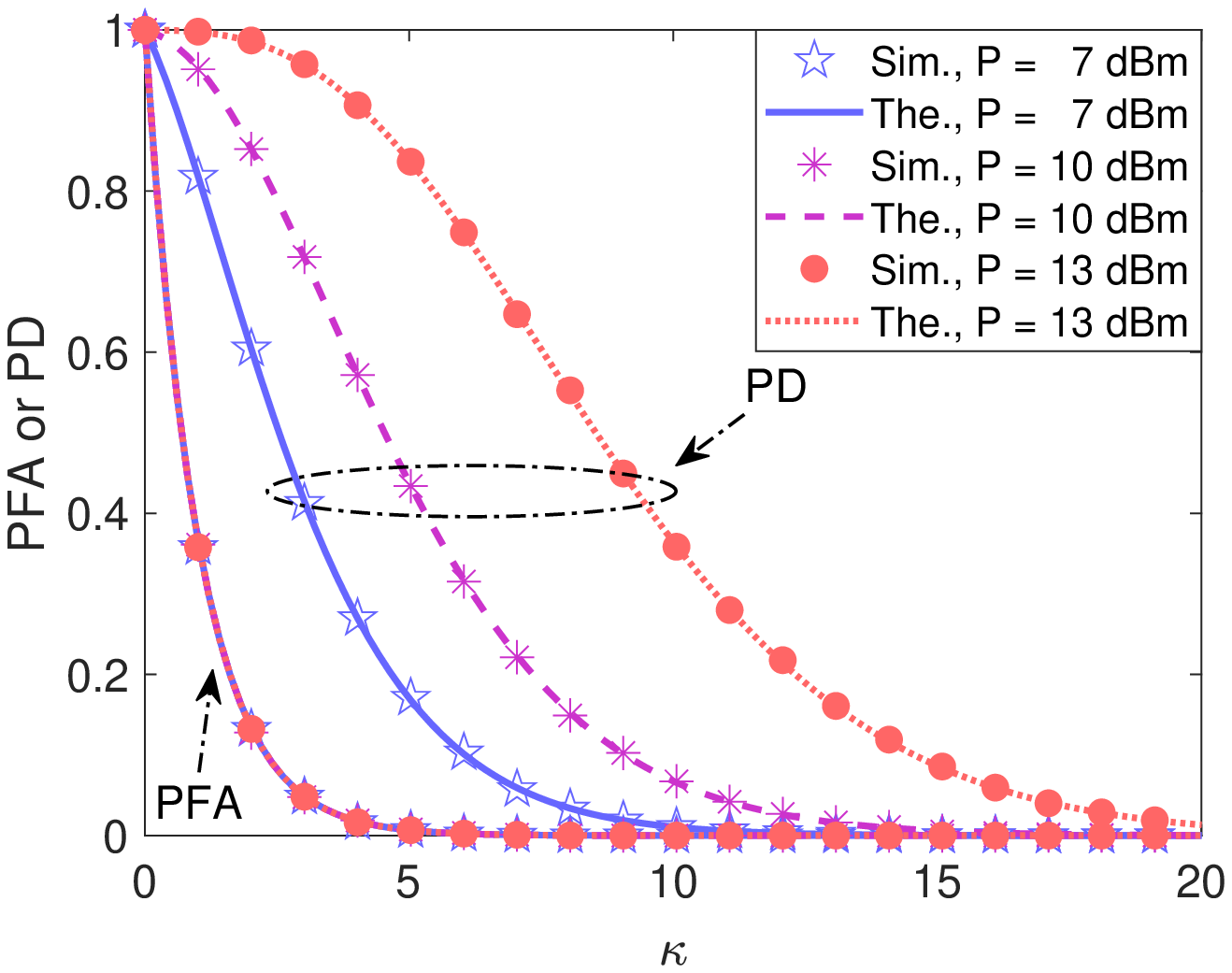}
\label{fig3b}}\\
\subfloat[$P_{\textrm{FA}}$ or $P_{\textrm{D}}$ versus $\kappa$ with estimated ${\mathbf{s}}_c$.]{\includegraphics[width=7.5cm]{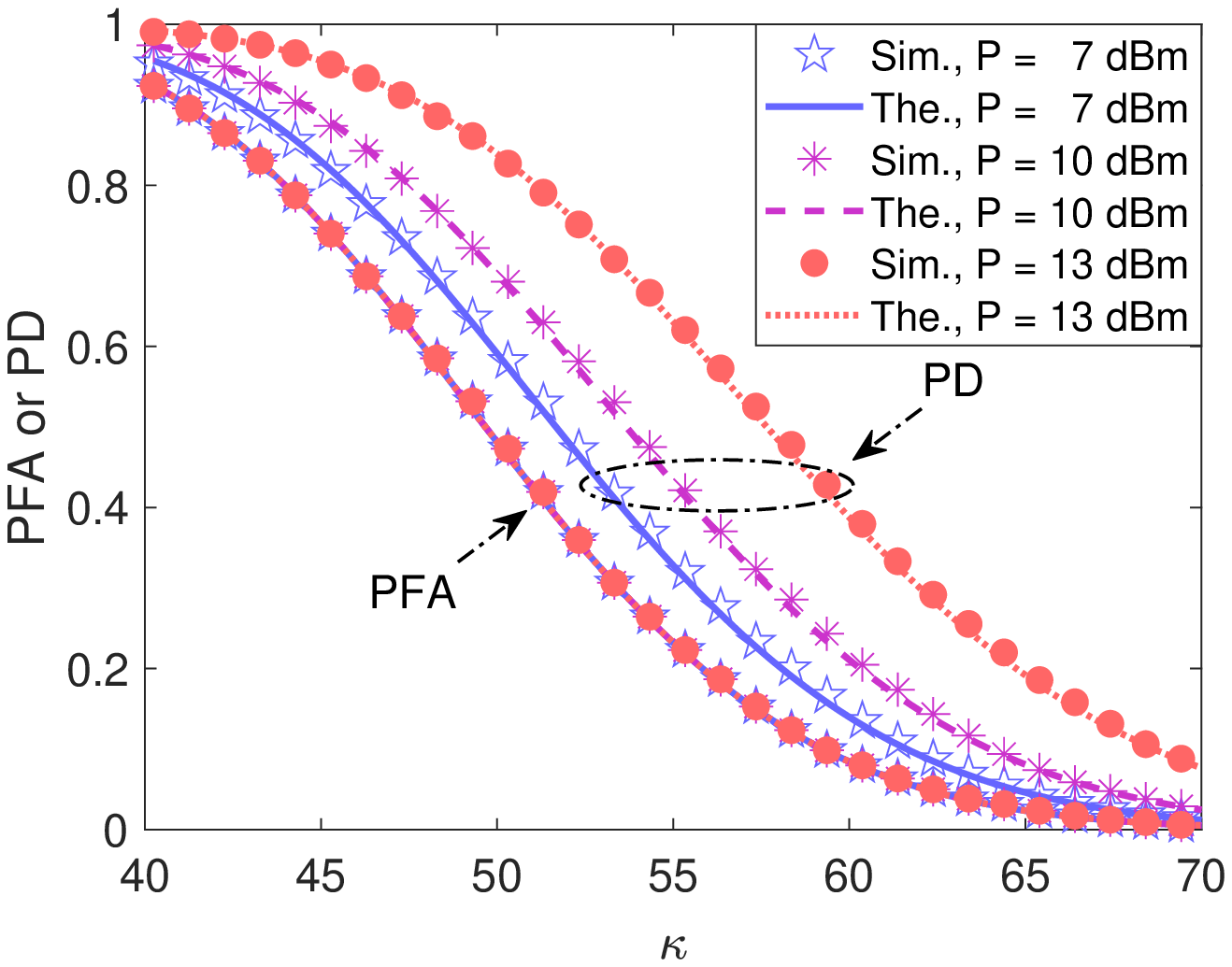}
\label{fig4a}}
\subfloat[$P_{\textrm{D}}$ versus $P_{\textrm{FA}}$.]{\includegraphics[width=7.5cm]{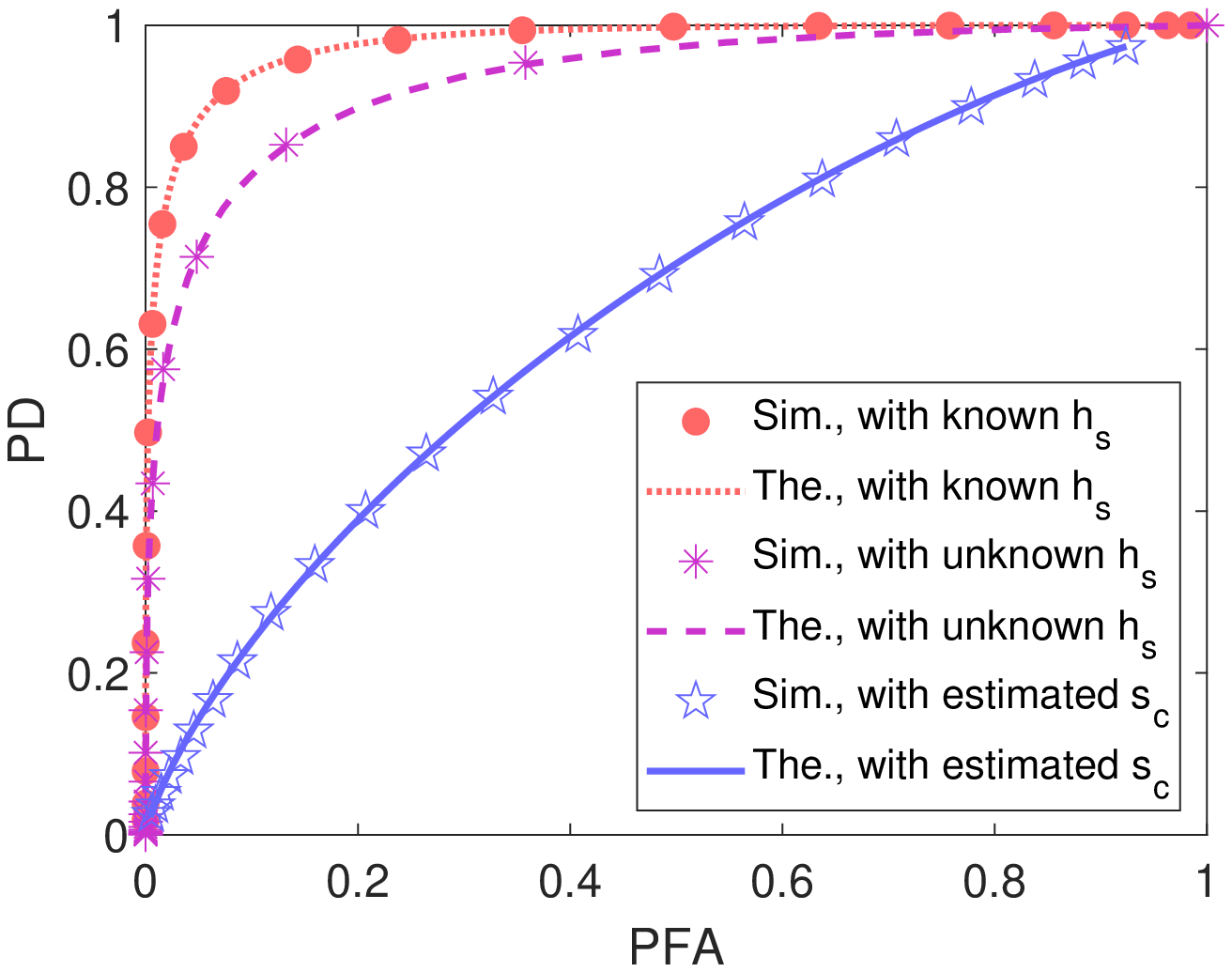}
\label{fig4b}}
\caption{Communication-assisted sensing scenario.}
\end{figure*}
\subsection{Validation of PFA and PD}
We first verify the accuracy of the derived theoretical expressions for PFA and PD by considering the communication-assisted sensing scenario with known $h_s$. We assume that $T = 50$ symbols are collected for performing coherent detection. The transmit power is set to $P = 7,\ 10,\ 13$ dBm. In order to obtain all possible values of PFA and PD from $0$ to $1$, the decision threshold, $\kappa$, is increased from $ - 2.5\lambda $ to $2.5\lambda $, where $\lambda = \frac{{P{{\left\| {{h_s}{\bf{s}}} \right\|}^2}}}{{\sigma _s^2}}$. In our simulations, we substitute $P = 10$ dBm when calculating $\lambda$ to maintain a unified decision threshold. The theoretical and simulated values of the PFA and the PD are plotted in Fig. \ref{fig3a}, where it can be observed that the theoretical values match perfectly with the simulated values of PFA and PD for all considered setups. As the transmit power increases, the detector gains more confidence to perform the target detection task, resulting in an increase in PD and a decrease in PFA. Next, Fig. \ref{fig3b} shows the PFA and PD under the communication-assisted sensing scenario with an unknown $h_s$, in which the decision threshold increases from $0$ to $ 5\lambda $ for obtaining all possible values of PFA and PD. Similarly, the analytical results perfectly predict the trends of PFA and PD. Note that the PFA remains constant with the increase of the transmit power $P$, due to the fact that the GLRT function in \eqref{eq17} is independent of the ISAC waveform under the null hypothesis, i.e., ${\mathcal{H}_0}$. By contrast, the PD increases with a growing value of $P$. For a given value of PFA, e.g., ${P_{\textrm{FA}}} = 0.1$, the PD increases from ${P_{\textrm{D}}} = 0.6$ to ${P_{\textrm{D}}} = 0.99$ when quadrupling the transmit power.

\begin{figure*}[!t]
\centering
\subfloat[$P_{\textrm{FA}}$ or $P_{\textrm{D}}$ versus $\kappa$ with known $h_s$ ($P = 8$ dBm).]{\includegraphics[width=7.5cm]{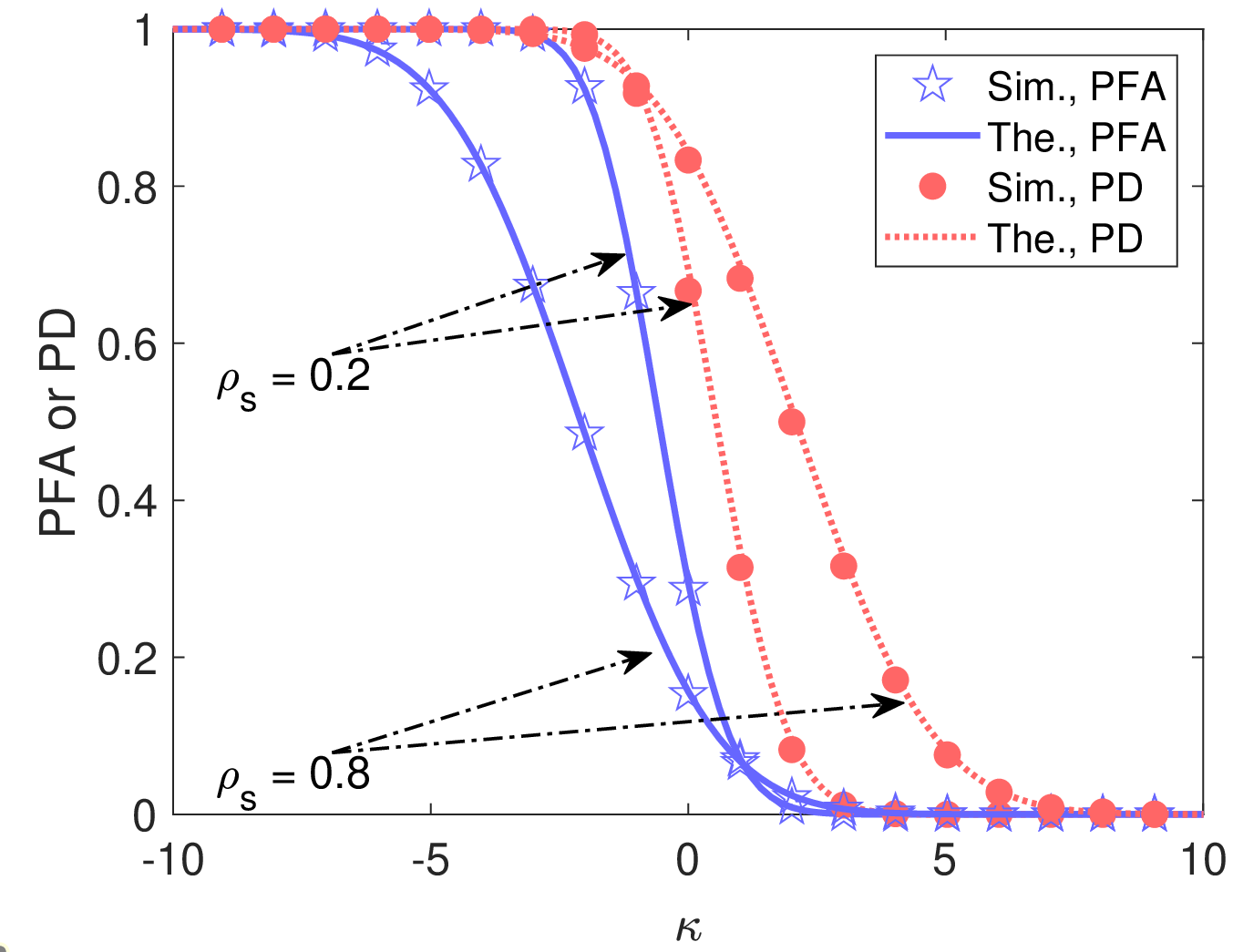}
\label{fig5a1}}
\subfloat[$P_{\textrm{FA}}$ or $P_{\textrm{D}}$ versus $\kappa$ with known $h_s$ ($P = 12$ dBm).]{\includegraphics[width=7.5cm]{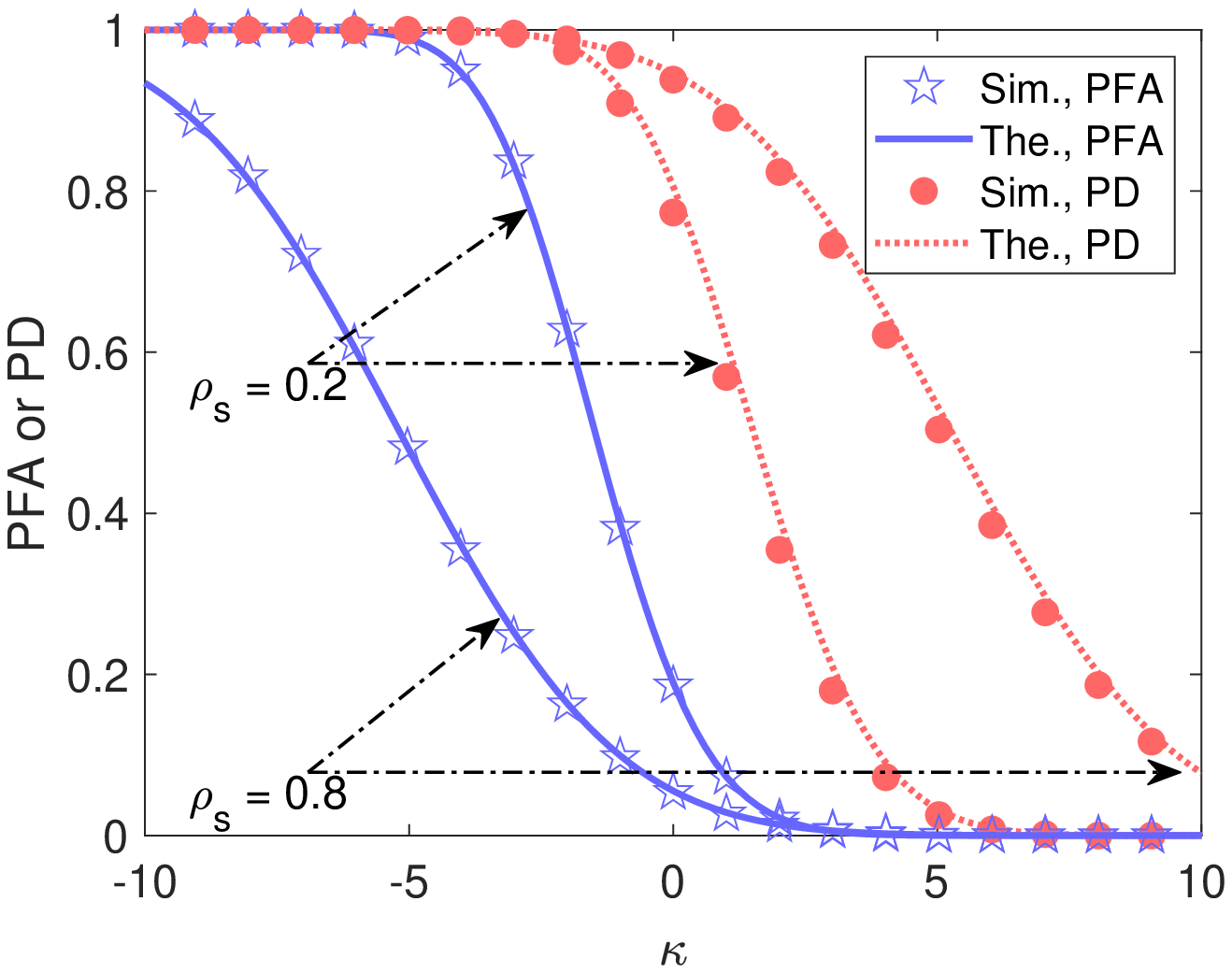}
\label{fig5a2}}\\
\subfloat[$P_{\textrm{FA}}$ or $P_{\textrm{D}}$ versus $\kappa$ with unknown $h_s$.]{\includegraphics[width=7.5cm]{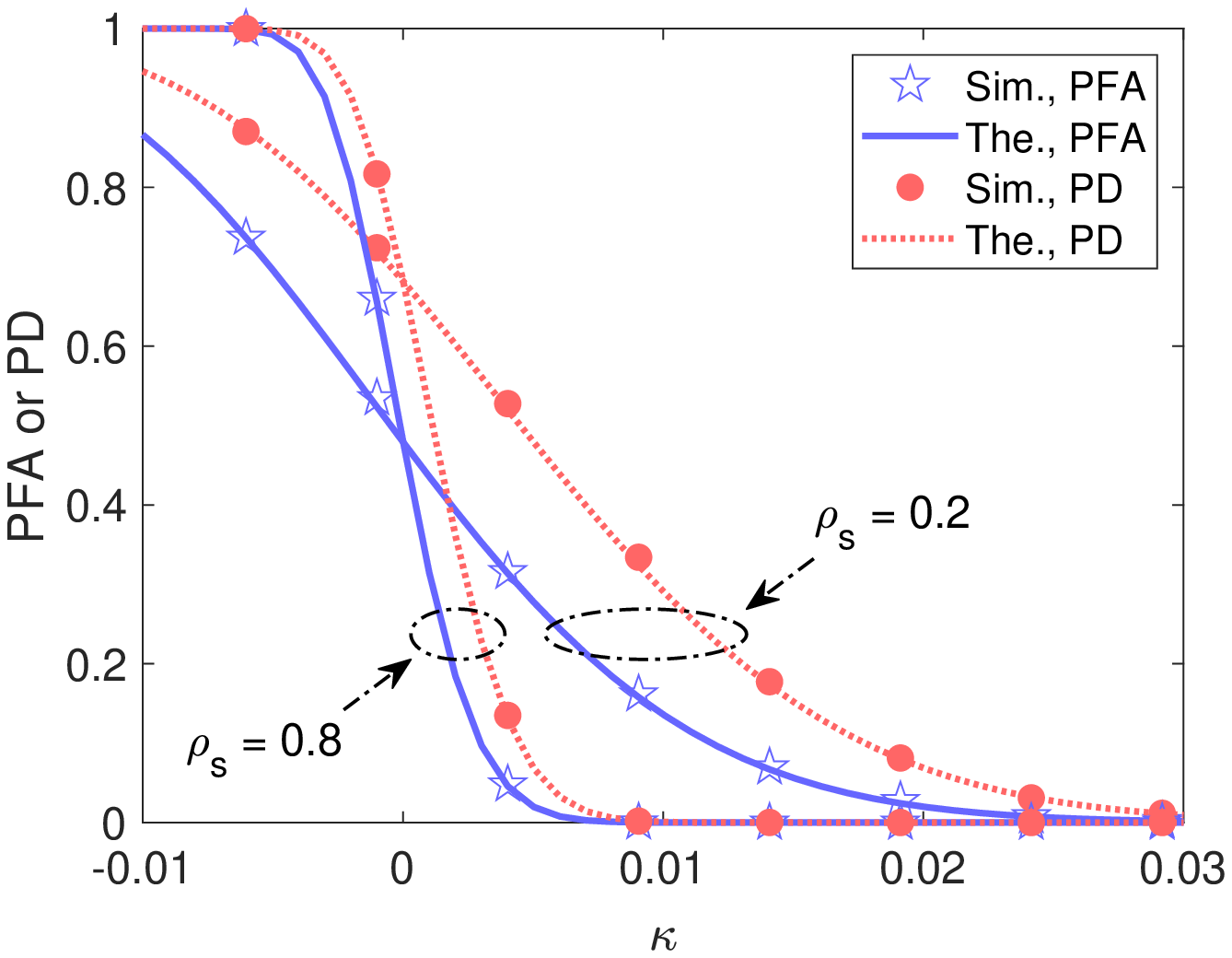}
\label{fig5b}}
\subfloat[$P_{\textrm{D}}$ versus $P_{\textrm{FA}}$.]{\includegraphics[width=7.5cm]{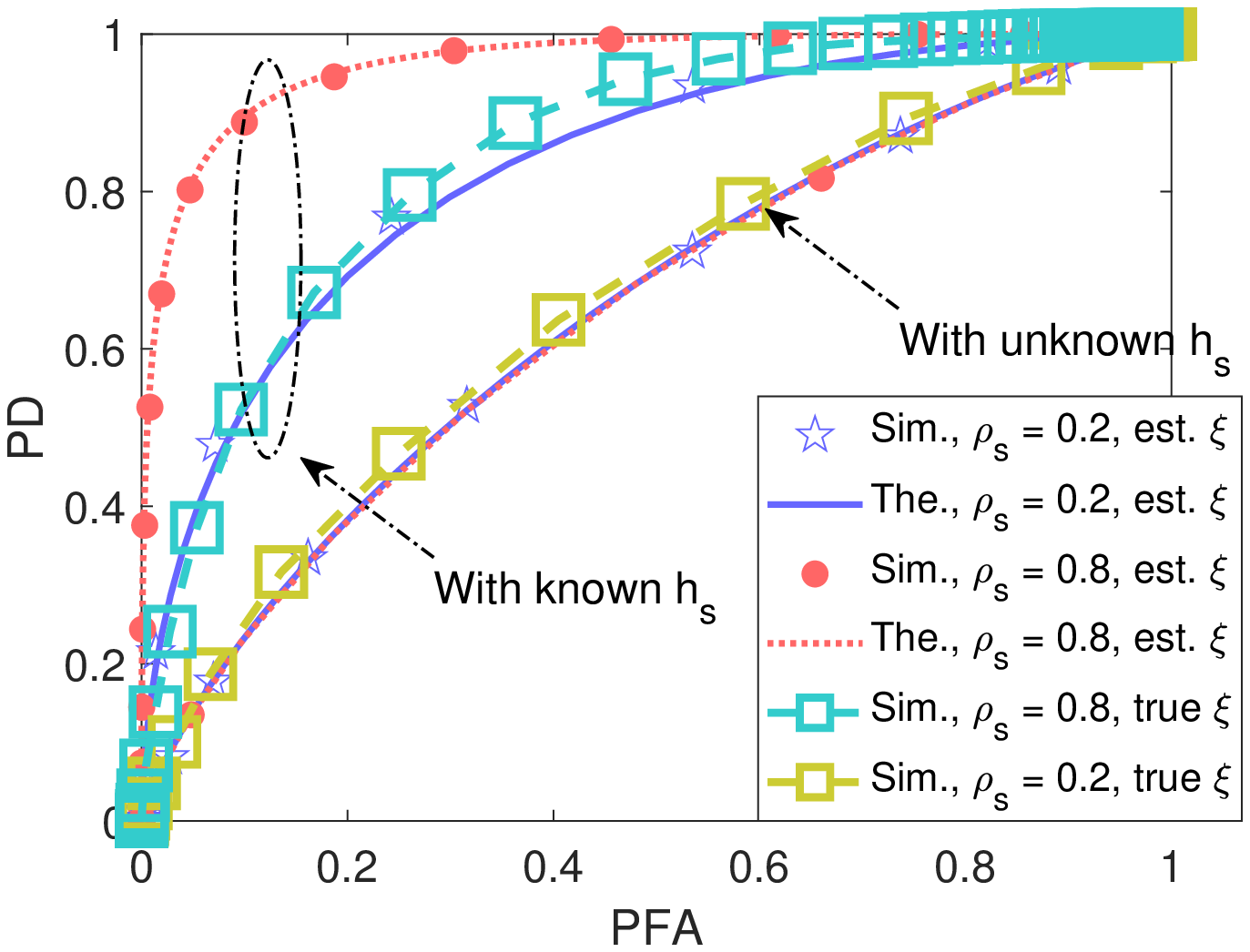}
\label{fig6a}}
\caption{Communication-interfered sensing scenario.}
\end{figure*}

Fig. \ref{fig4a} shows the PFA and PD of the communication-assisted sensing scenario, where the communication signal is estimated by using \eqref{eq21}. Following the same philosophy, the decision threshold increases from $ 10\lambda $ to $ 17.5\lambda $. The simulated PFA and PD are consistent with our previous theoretical analysis. Compared with the results in Figs. \ref{fig3a} and \ref{fig3b}, we observe that the PD curve in this scenario is closer to the PFA curve under the same transmit power, which implies the poor performance of the energy detector due to the lack of waveform information. To make a more intuitive comparison, the PD versus PFA under the above three cases are plotted in Fig. \ref{fig4b}, from which one could readily observe that the coherent detector, i.e., \eqref{eq12}, and the energy detector, i.e., \eqref{eq22}, serves as upper and lower bounds for the communication-assisted sensing scenario, respectively, since the former takes the full advantage of the communication waveform, while the latter exploits it the least. The communication-assisted sensing with unknown $h_s$ suffers from a moderate performance penalty compared to that having known $h_s$. Specifically, for a target PFA value of $P_{\textrm{FA}} = 0.1$, the PD in the communication-assisted sensing scenario is upper bounded by $0.95$, which reduces to $0.8$ without the prior information of $h_s$.

\begin{figure}[!t]
\centering
\subfloat[With known $h_s$ (\emph{Case I}, upper bound).]{\includegraphics[width=7.5cm]{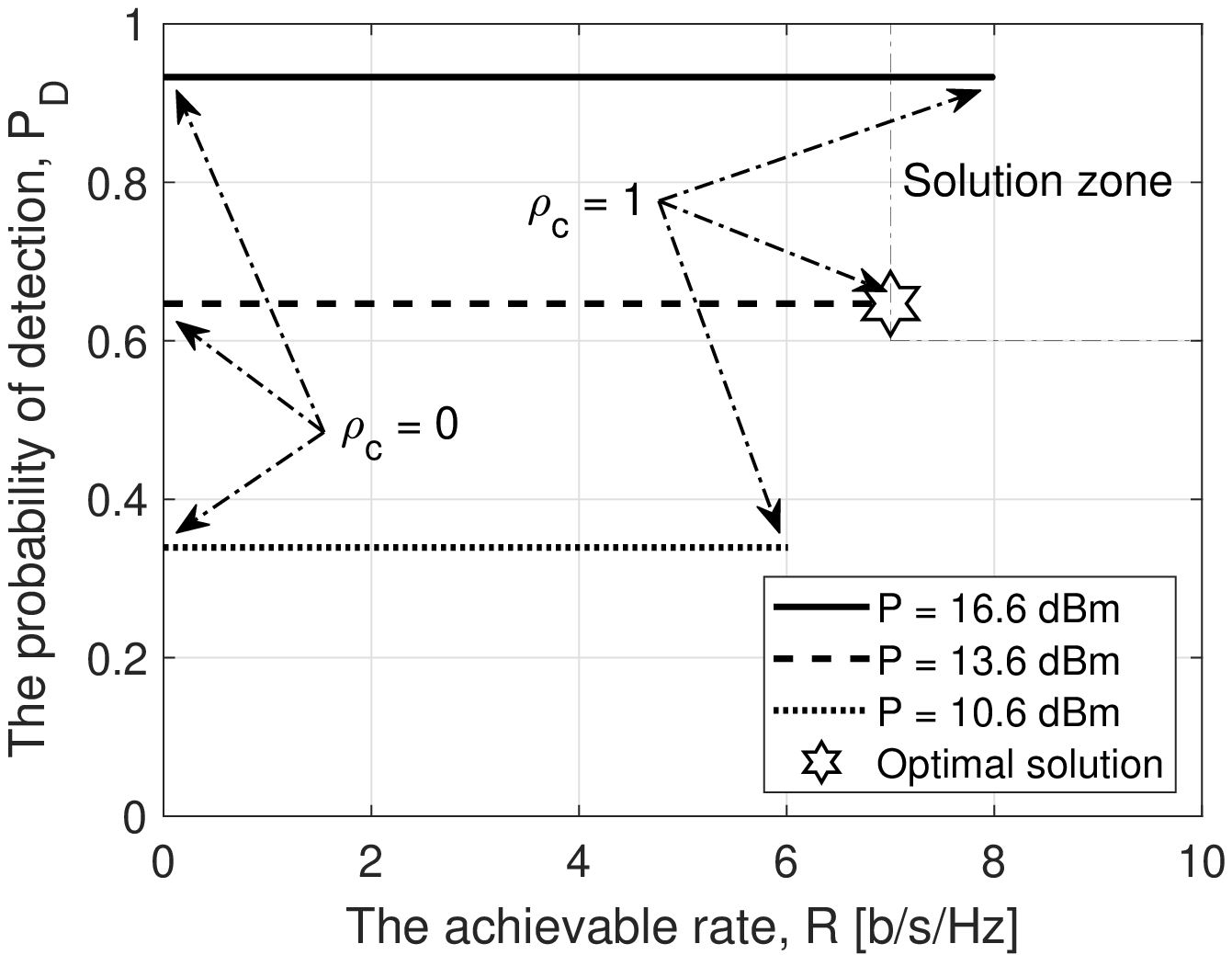}
\label{fig6b}}\\
\subfloat[With unknown $h_s$ (\emph{Case II}, upper bound).]{\includegraphics[width=7.5cm]{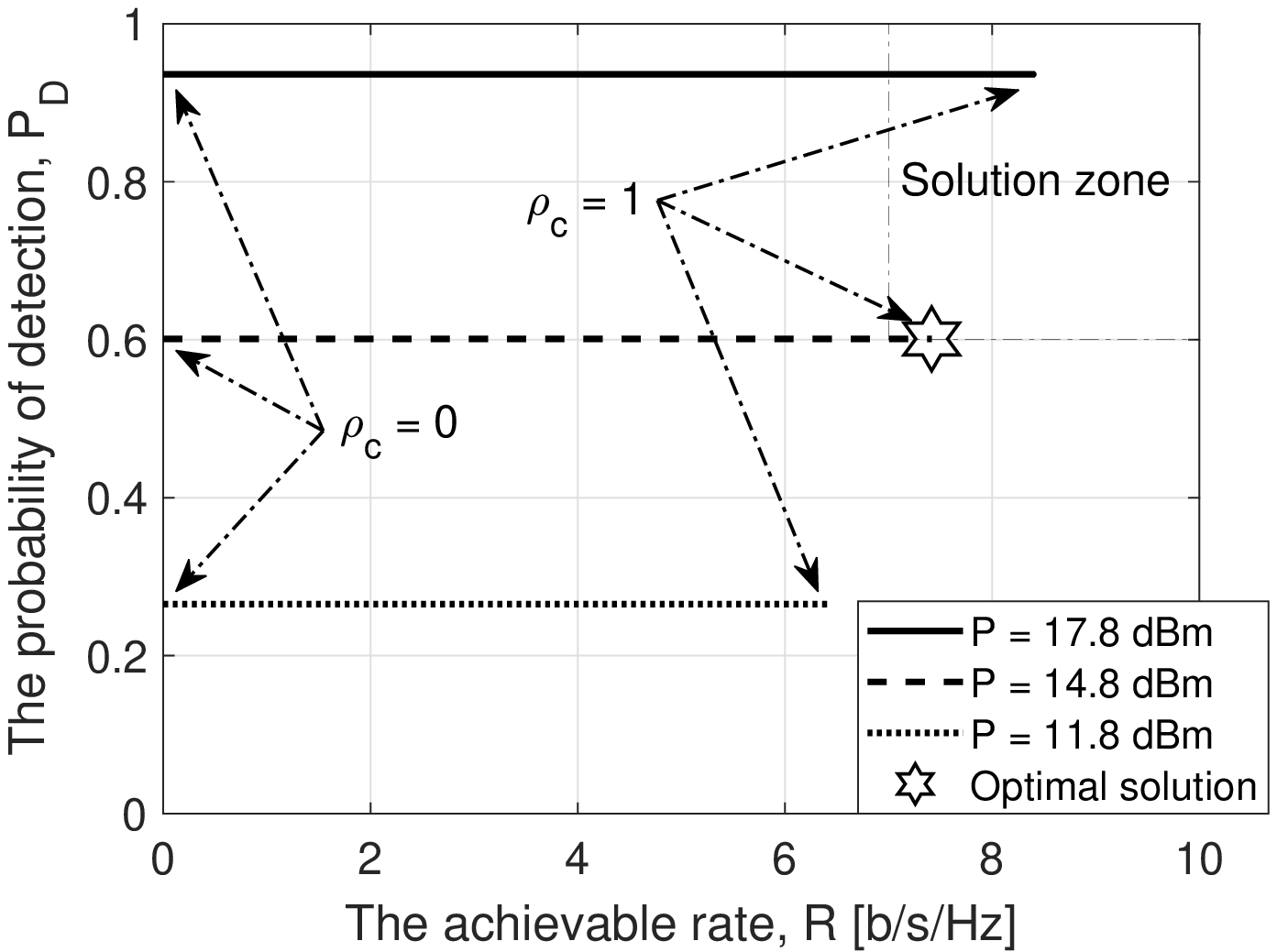}
\label{fig7a}}\\
\subfloat[With estimated ${\mathbf{s}_c}$ (\emph{Cases I and II}, lower bound).]{\includegraphics[width=7.5cm]{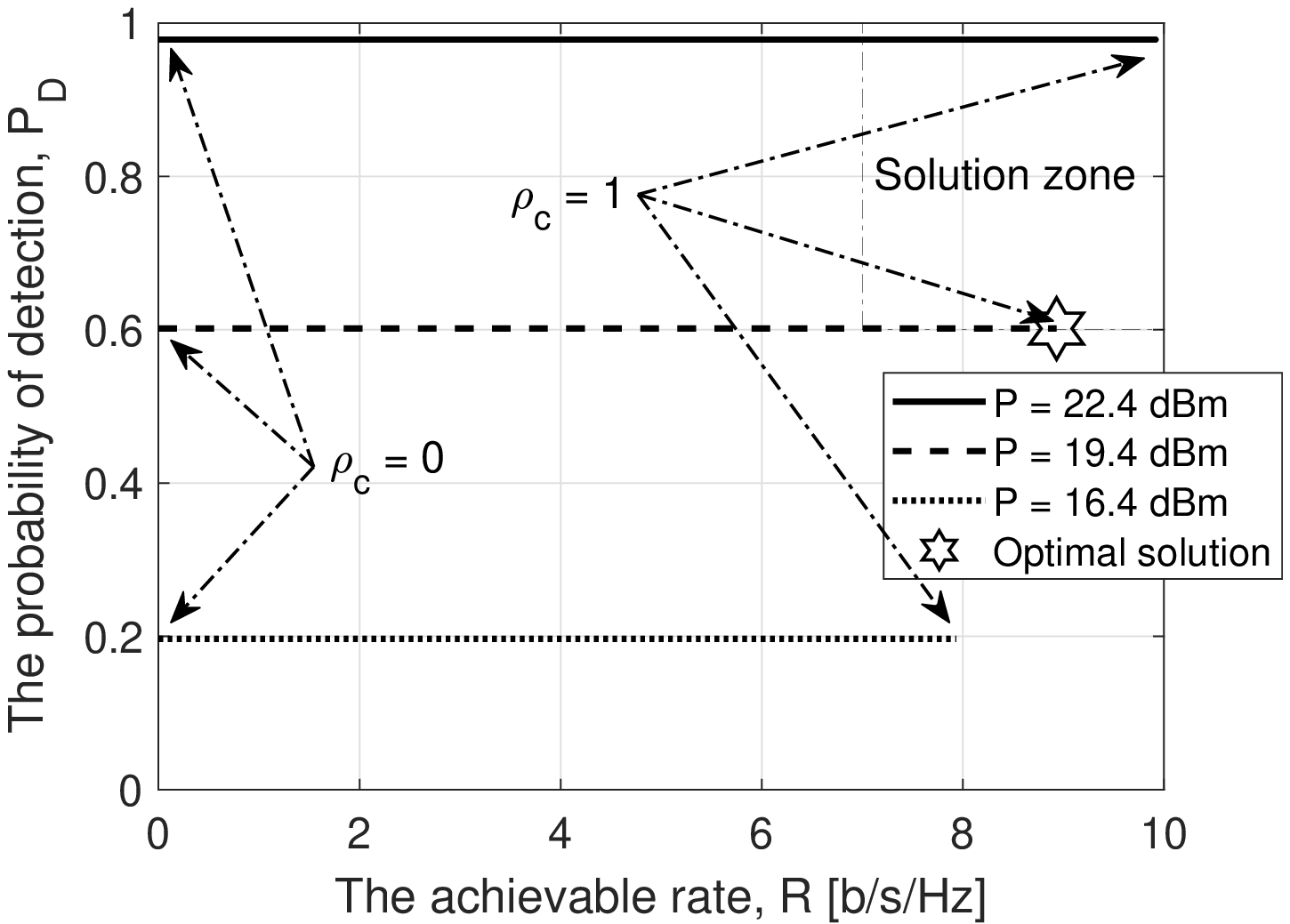}
\label{fig7b}}
\caption{$P_{\textrm{D}}$ versus $R$ for the ISAC scenario of sensing-free communication and communication-assisted sensing.}
\end{figure}
\begin{figure}[!t]
\centering
\subfloat[With known $h_s$ (\emph{Case III}).]{\includegraphics[width=7.5cm]{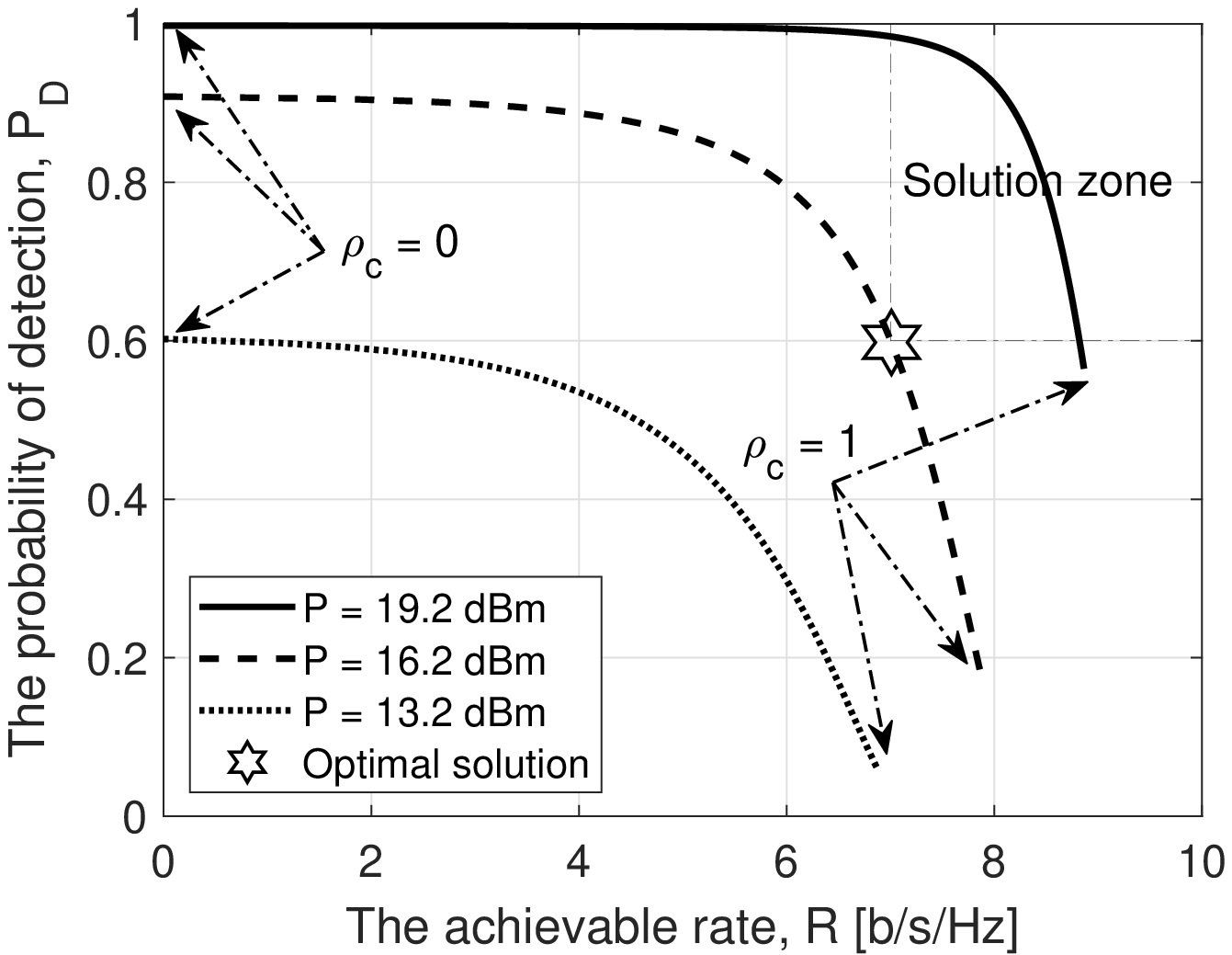}
\label{fig8a}}\\
\subfloat[With unknown $h_s$ (\emph{Case IV}).]{\includegraphics[width=7.5cm]{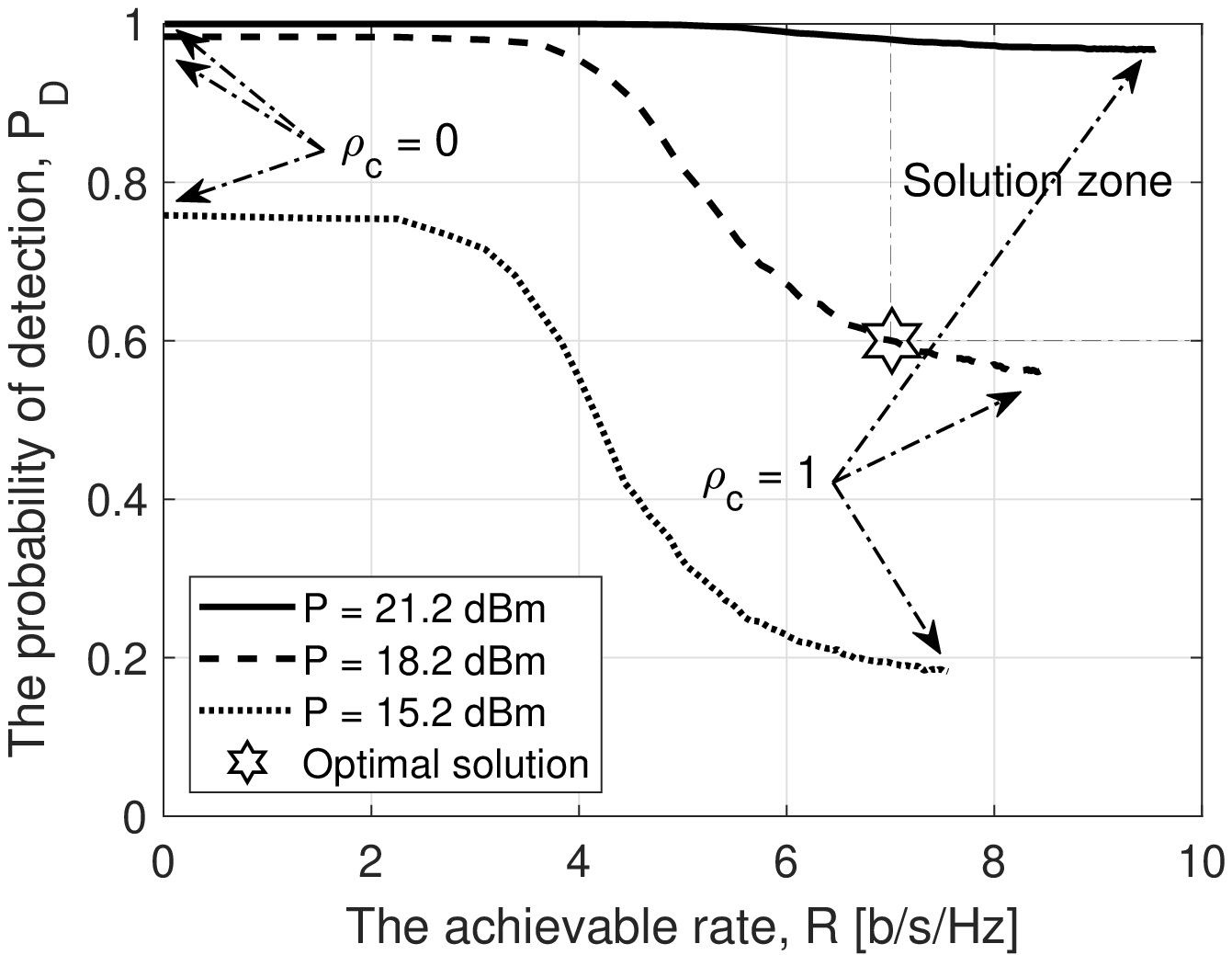}
\label{fig8b}}
\caption{$P_{\textrm{D}}$ versus $R$ for the ISAC scenario of sensing-free communication and communication-interfered sensing.}
\end{figure}
Next, we consider the communication-interfered scenario where the communication signal is unknown and acts as interference at the SR. The decision threshold increases from $ -0.25\lambda $ to $0.25\lambda $, while the normalized power coefficient allocated for sensing is set to ${\rho _s} = 0.8,\ 0.2$. The simulation results considering the transmit power of $P = 8$ dBm and $P = 12$ dBm are shown in Figs. \ref{fig5a1} and \ref{fig5a2}, respectively. It is demonstrated that PD gains improvement with the increase of ${\rho _s}$ and $P$, since more power for sensing is allocated and the interference is reduced. As ${\rho _s}\rightarrow 1$, the interference-free case in Fig. \ref{fig3a} serves as the upper bound of the communication-interfered scenario, in which all the power is utilized for sensing and the interference caused by the communication signal is minimized. Fig. \ref{fig5b} shows the PFA and PD under the communication-interfered scenario with unknown $h_s$. For the sake of brevity, we only consider the case with the transmit power of $P = 10$ dBm. In order to verify the accuracy of the derived lower bound \eqref{eq38}, we assume that the second-order statistics of the channel coefficient are known by the SR. The quadrature order for calculating \eqref{eq38} is set to $N = 1000$. Note that the theoretical analysis perfectly matches the simulated values of PFA and PD. With the increase of the sensing power coefficient, i.e., $\rho_{s}$, the PD under the same value of PFA improves moderately due to more sensing power collected and less communication interference incurred. In order to make an intuitive comparison, Fig. \ref{fig6a} plots the PD versus the PFA under the communication-interfered scenario. Note that without prior knowledge of $h_s$, the PD is reduced from $0.9$ to $0.2$ at the target PFA value of $0.1$. As the power coefficient for sensing increases from $0.2$ to $0.8$, the PD improves from $0.7$ to $0.95$ for the PFA target of $0.2$. For the sake of elaboration, we also plot the actual PD versus PFA curve by substituting the true value of $ \xi $ into the logarithmic GLRT function. Note that the derived PD in \eqref{eq38} serves as a lower bound of this case, due to the statistical approximation in our analysis. When considering the case with a sensing power coefficient of $\rho_s = 0.2$, the derived PD is a tight lower bound.

\subsection{Tradeoff Analysis between Sensing and Communication Functionalities}
Next, we will examine the tradeoff between the PD and the achievable rate in the context of an ISAC scenario. Specifically, $T = 20$ symbols are collected for performing target detection. The minimum requirement for the achievable rate at the CU is set to $R_{\min} = 7$ b/s/Hz, while the minimum PD requirement at the SR is $P_{\textrm{D},\min} = 0.6$ under the target PFA of $P_{\textrm{FA},\delta} = 0.01$. Fig. \ref{fig6b} illustrates the PD vs. achievable rate tradeoff for the ISAC scenario, considering sensing-free communication and communication-assisted sensing with known $h_s$, i.e., \emph{Case I}. Specifically, different operating boundaries represent the simulated $P_{\textrm{D}}$ vs. $R$ curve under different values of transmit power, while the hexagram represents the optimal power allocation solution obtained from our previous theoretical analysis. Note that in this case, the PD is independent of the power allocation coefficient $\rho_{c}$ due to the fact that the communication waveform is always known at the SR and can thus be employed for performing coherent detection. As a consequence, there is no doubt that the optimal power allocation solution occurs when $\rho_{c} = 1$. The theoretical value of the minimum transmit power also matches the simulated one, which is obtained by utilizing the bisection method and equals $P_{\min} = 13.6$ dBm in this case. For the sake of illustration, we also plot the results for the transmit power of $P_{\min}\pm 3$ dBm. It is evident that transmitting less power fails to meet both the achievable rate and PD requirements, while increasing power leads to reduced energy efficiency.

\begin{figure}[!t]
\centering
\subfloat[With known $h_s$ (\emph{Case V}, upper bound).]{\includegraphics[width=7.5cm]{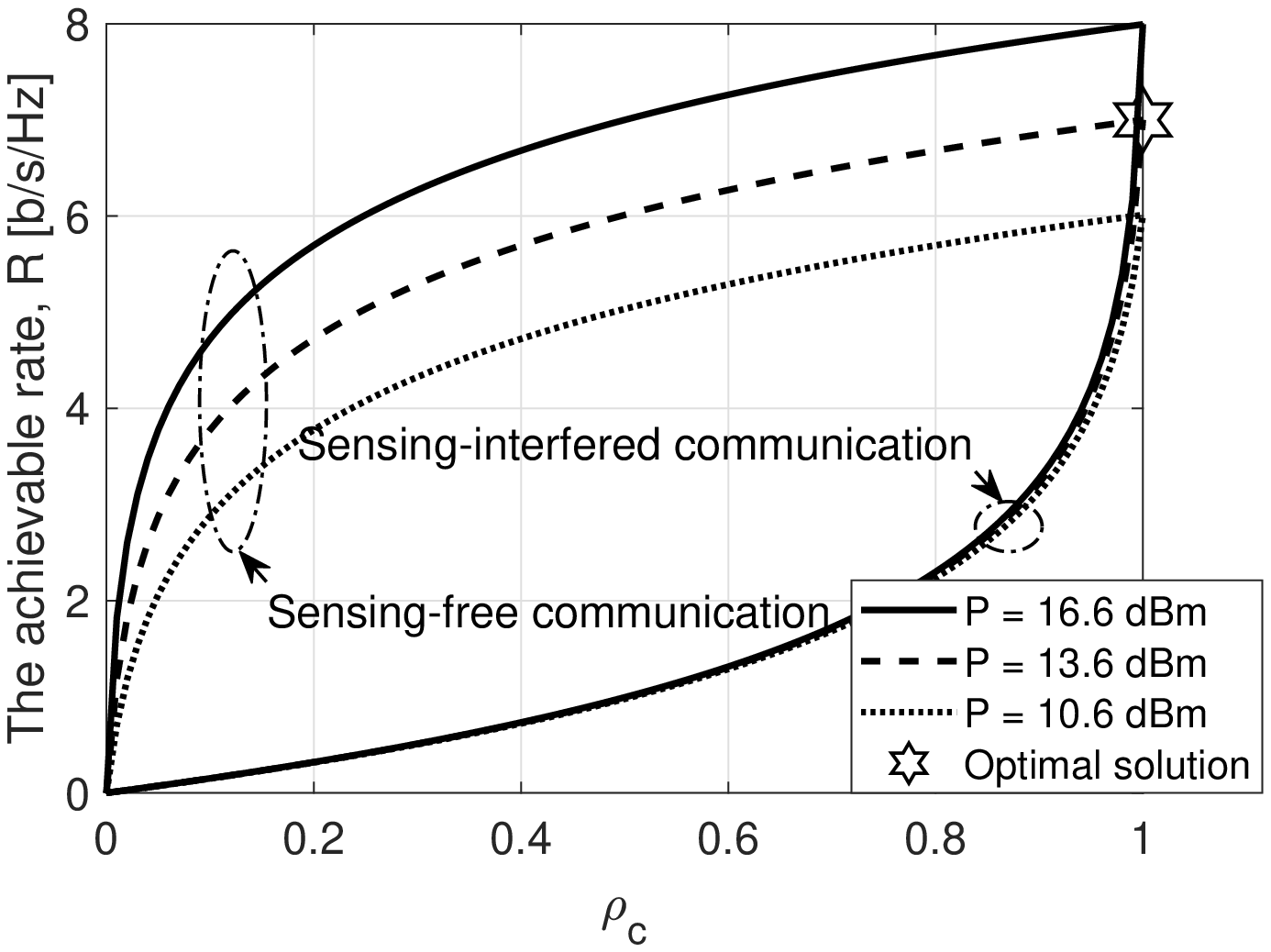}
\label{fig9a}}\\
\subfloat[With unknown $h_s$ (\emph{Case VI}, upper bound).]{\includegraphics[width=7.5cm]{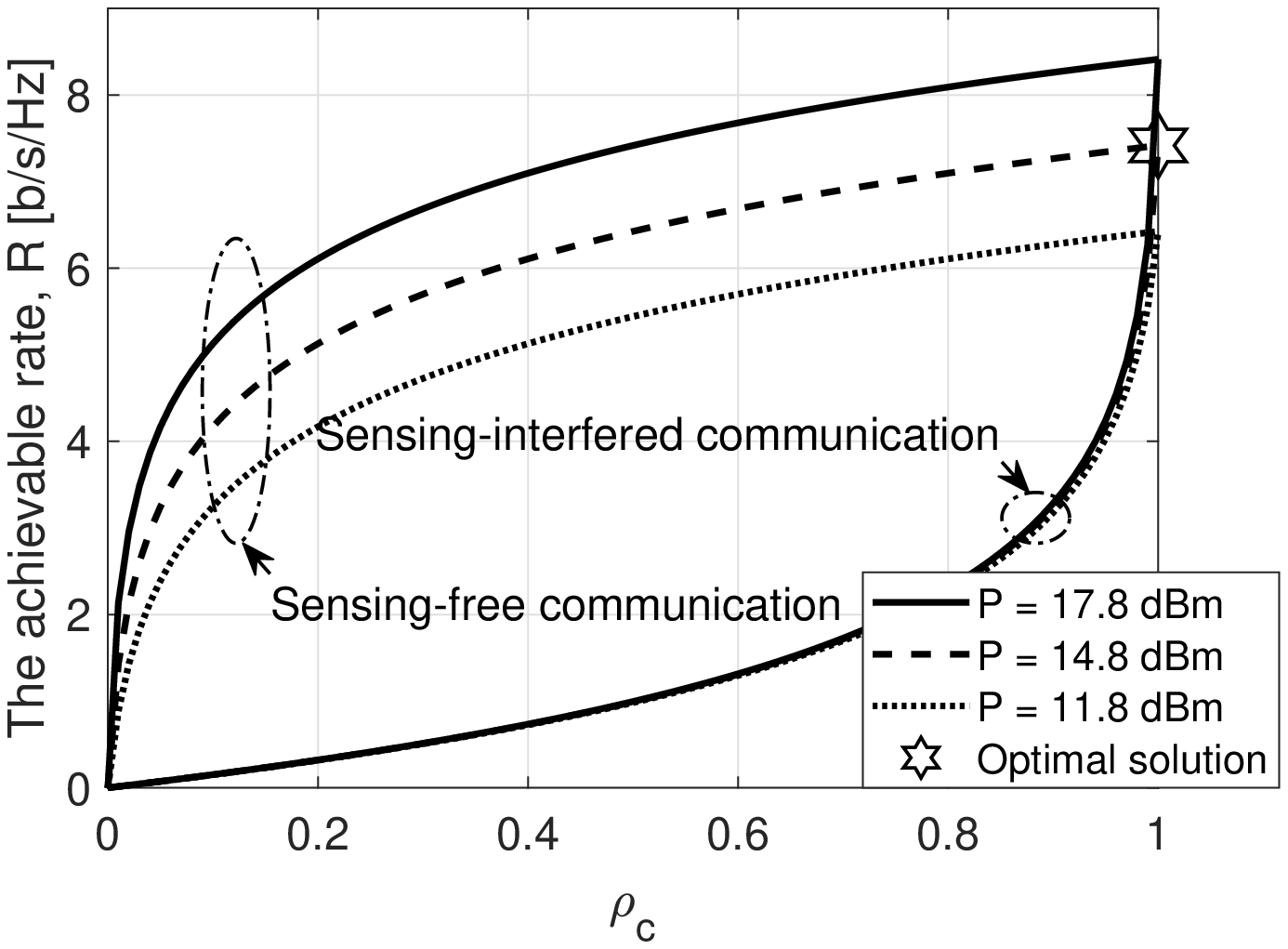}
\label{fig9b}}\\
\subfloat[With estimated ${\mathbf{s}}_c$ (\emph{Cases V \& VI}, lower bound).]{\includegraphics[width=7.5cm]{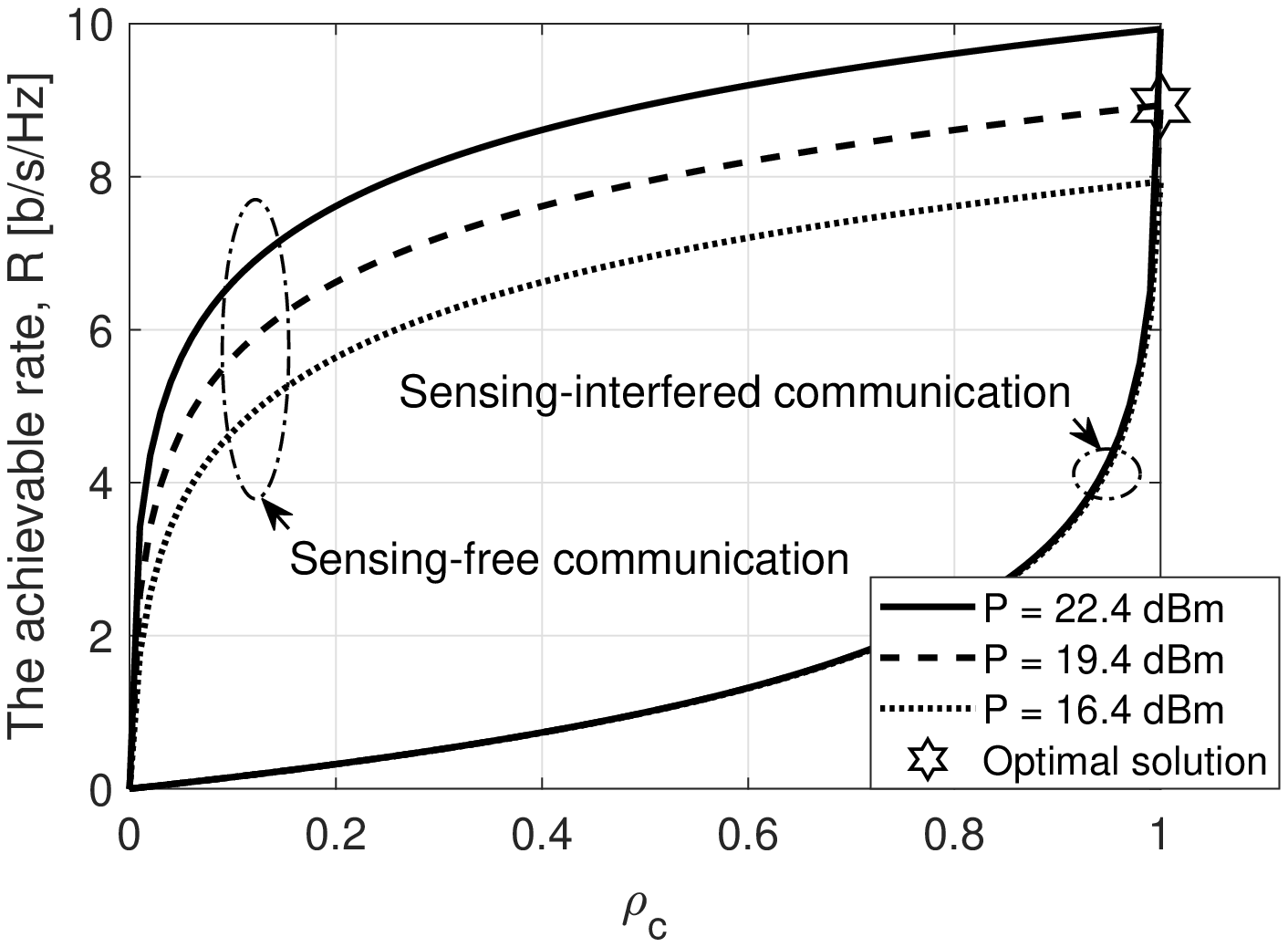}
\label{fig10a}}
\caption{$R$ versus $\rho_{c}$ for the ISAC scenario of sensing-interfered communication and communication-assisted sensing.}
\end{figure}
\begin{figure}[!t]
\centering
\subfloat[With known $h_s$ (\emph{Case VII}).]{\includegraphics[width=7.5cm]{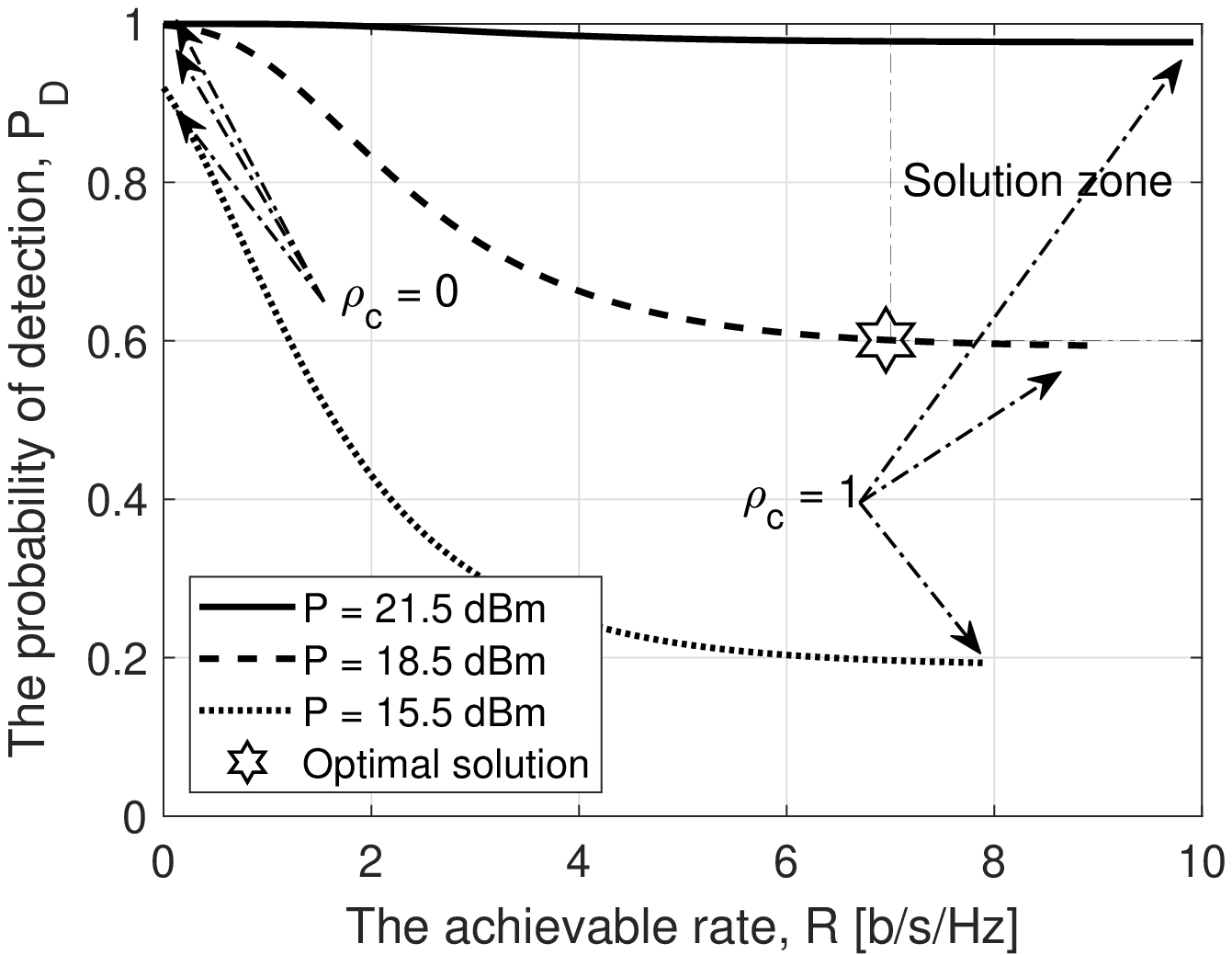}
\label{fig10b}}\\
\subfloat[With unknown $h_s$ (\emph{Case VIII}).]{\includegraphics[width=7.5cm]{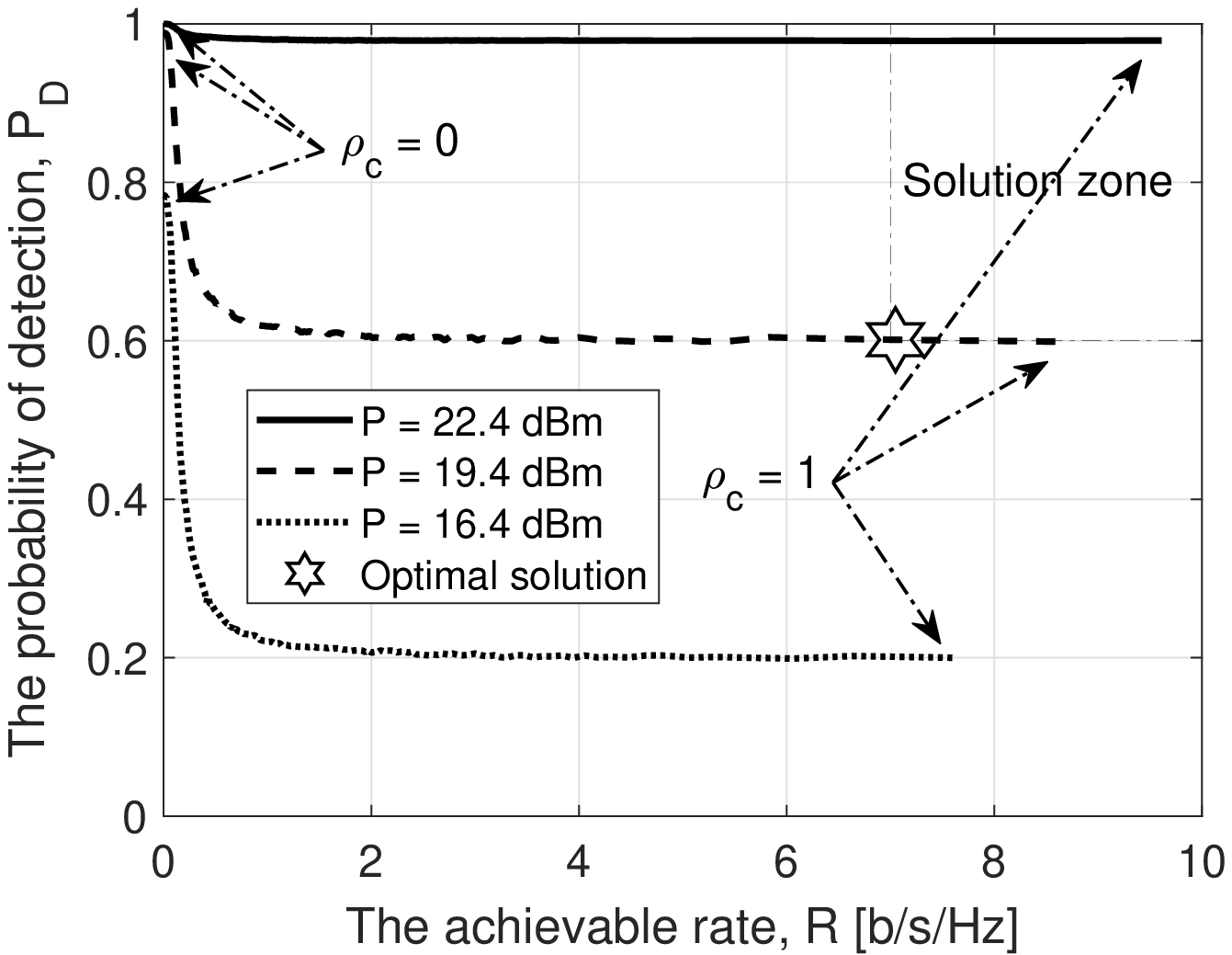}
\label{fig11a}}
\caption{$P_{\textrm{D}}$ versus $R$ for the ISAC scenario of sensing-interfered communication and communication-interfered sensing.}
\end{figure}
Furthermore, Fig. \ref{fig7a} plots the ISAC scenario of sensing-free communication and communication-assisted sensing with an unknown $h_s$, i.e., \emph{Case II}. The results are similar to those in Fig. \ref{fig6b}. The optimal power allocation solution is achieved at the point where $\rho_{c} = 1$, but the minimum transmit power required at the ISAC-BS increases from $13.6$ dBm to $14.8$ dBm due to the imperfect coherent detector. In Fig. \ref{fig7b}, we examine the communication-assisted sensing case with the estimated ${\mathbf{s}}_c$. One could obtain similar results to Fig. \ref{fig7a}. However, the required transmit power increases to $19.4$ dBm due to the lack of the communication waveform, which nearly quadruples the transmit power compared to \emph{Case I}. Additionally, one could observe from Figs. \ref{fig7a} and \ref{fig7b} that the theoretical transmit power and the optimal power allocation solution match the simulation results accurately. By increasing or reducing the transmit power at the ISAC-BS, it is shown that less power will lead to an infeasible solution while more power will result in energy waste.

Fig. \ref{fig8a} shows the results considering the ISAC scenario of sensing-free communication and communication-interfered sensing with known $h_s$ (\emph{Case III}). Note that since the communication signal acts as interference during the sensing procedure, there exists an intrinsic tradeoff in this case. It is evident that the optimal power allocation solution occurs at the edge of the $P_{\textrm{D}}$-$R$ curve. Therefore, the minimum power coefficient allocated for communication satisfying the achievable rate requirement is $\rho_{c}P = R_{\min}$. The optimal power allocation coefficient in this case is $0.55$, which matches the theoretical results. Furthermore, we consider the ISAC scenario of sensing-free communication and communication-interfered sensing with unknown $h_s$ (\emph{Case IV}). Due to the communication interference and the lack of amplitude information, the PD degrades severely with the increase of $\rho_{c}$, necessitating more power to reach the solution zone. As depicted in Fig. \ref{fig8b}, the minimum transmit power required to meet the communication and sensing requirements is $P_{\min} = 18.2$ dBm, which results in $2$ dB performance erosion compared to \emph{Case III} in Fig. \ref{fig8a}. The optimal power allocation coefficient in this case is $\rho_c = 0.34$.

Next, we consider the ISAC scenario of sensing-interfered communication and communication-assisted sensing. Note that the optimal power allocation solution and the $P_{\textrm{D}}$-$R$ curve in this case are the same as those in the sensing-free communication scenario. Hence, we plot the $R$-$\rho_c$ curve to characterize the difference under different power allocation coefficients. The communication-assisted scenario with known $h_s$, with unknown $h_s$, with estimated ${\mathbf{s}}_c$ are shown in Figs. \ref{fig9a}, \ref{fig9b}, \ref{fig10a}, respectively. Note that the sensing-free communication scenario attains a higher achievable rate than the sensing-interfered communication scenario owing to the interference caused by the sensing signal in the latter. For example, when considering the transmit power of $P = 16.6$ dBm with a power allocation coefficient of $\rho_c = 0.1$, the sensing-free communication scenario gains almost a ten-fold rate improvement compared to the sensing-interfered counterpart. Nevertheless, as the power allocation coefficient for communication increases, the sensing interference in the sensing-interfered communication scenarios gradually wears off. As a consequence, both two cases ultimately achieve the identical achievable rate, which is also the optimal power allocation solution. The same results can be observed in Figs. \ref{fig9b} and \ref{fig10a}.

Fig. \ref{fig10b} shows the performance tradeoff in the ISAC scenario of sensing-interfered communication and communication-interfered sensing with known $h_s$ at the SR. In this case, the communication signal plays the role of interference, resulting in a lower PD and achievable rate compared to that depicted in Fig. \ref{fig8a}. Consequently, more transmit power is demanded to meet the achievable rate requirement. Note that the minimum transmit power in this case is $18.5$ dBm, which is $2$ dB higher than the sensing-free communication counterpart. The optimal power allocation coefficient in this case is obtained by $\rho_{c} = 0.99$, which allocates almost the total available power for communication. Observing from Fig. \ref{fig10b} that the simulation result closely matches the optimal power allocation solution. Furthermore, the ISAC scenario of sensing-interfered communication and communication-interfered sensing with unknown $h_s$ are shown in Fig. \ref{fig11a}. In order to capture the intrinsic tradeoff, we consider the realistic case where $\xi$ is estimated instead of using a fixed value in the lower bound. Without the knowledge of the waveform and channel coefficient at the SR, the PD deteriorates severely as $\rho_c$ increases. The optimal power allocation coefficient is $\rho_c=0.995$, which means that achieving the preset achievable rate target is more challenging than meeting the PD target. As a result, almost all available power is allocated for communication. Only a small amount of power allocated for sensing with strong interference caused by communication signals is still sufficient to meet the PD target of $P_{\textrm{D},\min}=0.6$.

\begin{figure}[!t]
\centering
\includegraphics[width=7.5cm]{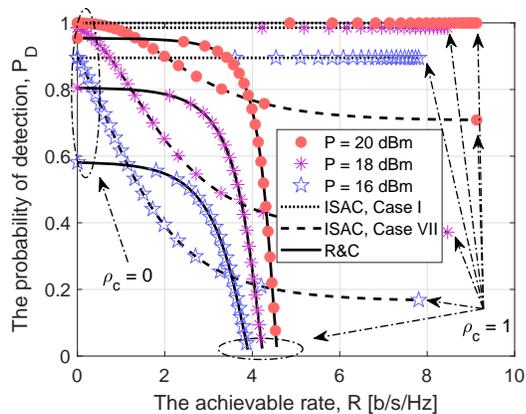}
\caption{The performance comparison of the ISAC system and the conventional R\&C system.\label{fig11b}}
\end{figure}
Finally, we compare the performance of the conventional R\&C coexistence scenario with the advanced ISAC system. In the R\&C system, we consider the time orthogonal mode for sensing and communication, which means that only half of the symbols are collected for sensing. Moreover, these two functionalities in the R\&C system operate competitively with regards to the power utilization, where the portion of power allocated for radar sensing and communication in the R\&C coexistence system are also denoted by $\rho_s$ and $\rho_c$, respectively. As a result, the PD in \eqref{eq15} and PFA in \eqref{eq14} characterize the theoretical performance of the R\&C coexistence system, upon replacing the total power $P$ with the sensing power $\rho_{s} P$ and the symbol length $T$ with $T/2$. The simulation results are shown in Fig. \ref{fig11b}, where we consider two ISAC systems: \emph{Case I} and \emph{Case VII}. Note that for the favorable ISAC scenario in which these two functionalities assist each other, ISAC outperforms the R\&C system doubtlessly. When considering the ISAC systems where two operations behave competitively in terms of resource utilization and suffer from mutual interference, ISAC performs better only at both ends of the $P_{\textrm{D}}$-$R$ curve, where one of the sensing and communication tasks is major. To be more specific, the sensing/communication interference at both ends is negligible, thus the ISAC system utilizing the total resource blocks achieves better performance than the R\&C system which only employs half. When both sensing and communication tasks impose stringent QoS requirements, the interference in the competitive-type ISAC system would deteriorate the other half's performance and thus both two operations suffer moderate performance erosion compared to the conventional R\&C system.

\section{Conclusions}\label{sec6}
This paper constructed a comprehensive framework for theoretically analyzing the performance of ISAC systems. We derived closed-form expressions to investigate the PFA and the PD under the communication-assisted and communication-interfered sensing scenarios. Based on our analysis, we discussed the fundamental tradeoff between a pair of sensing and communication metrics: the PD and the achievable rate, by solving the formulated power allocation problem. We obtained the optimal power allocation solution under different cases and elaborated on the effects of the sensing or communication signal on the other half's functionality. Finally, extensive simulation results verified our theoretical analysis. It is demonstrated that when sensing and communication capabilities are operated collaboratively, they could achieve mutual gain from each other, whereas there exists an intrinsic tradeoff when they operate in a competitive manner. Our simulation results also verified the benefits of the ISAC system operating a collaborative sensing and communication mode over the conventional R\&C coexistence counterpart.

\bibliography{ref}
\bibliographystyle{IEEEtran}

\end{document}